\documentclass[fleqn,usenatbib]{mnras}

\usepackage{newtxtext,newtxmath}
\usepackage[T1]{fontenc}

\DeclareRobustCommand{\VAN}[3]{#2}
\let\VANthebibliography\thebibliography
\def\thebibliography{\DeclareRobustCommand{\VAN}[3]{##3}\VANthebibliography}

\usepackage{threeparttable}
\usepackage{graphicx}	% Including figure files
\usepackage{amsmath}	% Advanced maths commands
\usepackage[table]{xcolor}
\usepackage{hhline}
\title[HERA Upper Limits on the EoR using the Closure Phase]{Search for the Epoch of Reionisation with HERA: Upper Limits on the Closure Phase Delay Power Spectrum.}

\author[P.\ M.\ Keller et al.]{	Pascal M. Keller,$^{1}$\thanks{Email: pmk46@cam.ac.uk}
	Bojan  Nikolic,$^{1}$
	Nithyanandan  Thyagarajan,$^{2}$
	Chris L. Carilli,$^{3}$
\newauthor
	Gianni  Bernardi,$^{4,5,6}$
	Ntsikelelo  Charles,$^{5}$
	Landman Bester,$^{6,5}$
	Oleg~M. Smirnov,$^{5,6}$
\newauthor
	Nicholas S. Kern,$^{8}$
	Joshua S. Dillon,$^{7}$
	Bryna J. Hazelton,$^{9,10}$
	Miguel F. Morales,$^{9}$
\newauthor
	Daniel C. Jacobs,$^{11}$
	Aaron R. Parsons,$^{7}$
	Zara  Abdurashidova,$^{7}$
	Tyrone  Adams,$^{6}$
\newauthor
	James E. Aguirre,$^{12}$
	Paul  Alexander,$^{1}$
	Zaki S. Ali,$^{7}$
	Rushelle  Baartman,$^{6}$
	Yanga  Balfour,$^{6}$
\newauthor
	Adam P. Beardsley,$^{11,13,\dagger}$
	Tashalee S. Billings,$^{12}$
	Judd D. Bowman,$^{11}$
	Richard F. Bradley,$^{14}$
\newauthor
	Philip  Bull,$^{15,16}$
	Jacob  Burba,$^{15,17}$
	Steven  Carey,$^{1}$
	Carina  Cheng,$^{7}$
	David R. DeBoer,$^{18}$
\newauthor
	Eloy  de~Lera~Acedo,$^{1}$
	Matt  Dexter,$^{18}$
	Nico  Eksteen,$^{6}$
	John  Ely,$^{1}$
	Aaron  Ewall-Wice,$^{7,19}$
\newauthor
	Nicolas  Fagnoni,$^{1}$
	Randall  Fritz,$^{6}$
	Steven R. Furlanetto,$^{20}$
	Kingsley  Gale-Sides,$^{1}$
\newauthor
	Brian  Glendenning,$^{3}$
	Deepthi  Gorthi,$^{7}$
	Bradley  Greig,$^{21}$
	Jasper  Grobbelaar,$^{6}$
\newauthor
	Ziyaad  Halday,$^{6}$
	Jacqueline N. Hewitt,$^{22,8}$
	Jack  Hickish,$^{18}$
	Austin  Julius,$^{6}$
\newauthor
	MacCalvin  Kariseb,$^{6}$
	Joshua  Kerrigan,$^{17}$
	Piyanat  Kittiwisit,$^{16}$
	Saul A. Kohn,$^{12}$
\newauthor
	Matthew  Kolopanis,$^{11}$
	Adam  Lanman,$^{17}$
	Paul  La~Plante,$^{7,12}$
	Adrian  Liu,$^{23}$
\newauthor
	Anita  Loots,$^{6}$
	Yin-Zhe  Ma,$^{24}$
	David Harold~Edward MacMahon,$^{18}$
	Lourence  Malan,$^{6}$
\newauthor
	Cresshim  Malgas,$^{6}$
	Keith  Malgas,$^{6}$
	Bradley  Marero,$^{6}$
	Zachary E. Martinot,$^{12}$
\newauthor
	Andrei  Mesinger,$^{25}$
	Mathakane  Molewa,$^{6}$
	Tshegofalang  Mosiane,$^{6}$
	Steven G. Murray,$^{11}$
\newauthor
	Abraham R. Neben,$^{8}$
	Hans  Nuwegeld,$^{6}$
	Robert  Pascua,$^{7,23}$
	Nipanjana  Patra,$^{7}$
\newauthor
	Samantha  Pieterse,$^{6}$
	Jonathan C. Pober,$^{17}$
	Nima  Razavi-Ghods,$^{1}$
	James  Robnett,$^{3}$
\newauthor
	Kathryn  Rosie,$^{6}$
	Mario G. Santos,$^{6,16}$
	Peter  Sims,$^{23}$
	Craig  Smith,$^{6}$
	Hilton  Swarts,$^{6}$
\newauthor
	Pieter  Van~Wyngaarden,$^{6}$
	Peter K.~G. Williams,$^{26,27}$
	Haoxuan  Zheng,$^{8}$
 \newauthor
\\
$^{1}$ Cavendish Astrophysics, University of Cambridge, Cambridge, UK\\
$^{2}$ Commonwealth Scientific and Industrial Research Organisation (CSIRO), Space \& Astronomy, P. O. Box 1130, Bentley, WA 6102, Australia\\
$^{3}$ National Radio Astronomy Observatory, Socorro, NM 87801, USA\\
$^{4}$ INAF-Istituto di Radioastronomia, via Gobetti 101, 40129 Bologna, Italy\\
$^{5}$ Department of Physics and Electronics, Rhodes University, PO Box 94, Grahamstown, 6140, South Africa\\
$^{6}$ South African Radio Astronomy Observatory, Black River Park, 2 Fir Street, Observatory, Cape Town, 7925, South Africa\\
$^{7}$ Department of Astronomy, University of California, Berkeley, CA\\
$^{8}$ Department of Physics, Massachusetts Institute of Technology, Cambridge, MA\\
$^{9}$ Department of Physics, University of Washington, Seattle, WA\\
$^{10}$ eScience Institute, University of Washington, Seattle, WA\\
$^{11}$ School of Earth and Space Exploration, Arizona State University, Tempe, AZ\\
$^{12}$ Department of Physics and Astronomy, University of Pennsylvania, Philadelphia, PA\\
$^{13}$ Department of Physics, Winona State University, Winona, MN\\
$^{\dagger}$ NSF Astronomy and Astrophysics Postdoctoral Fellow\\
$^{14}$ National Radio Astronomy Observatory, Charlottesville, VA\\
$^{15}$ Jodrell Bank Centre for Astrophysics, University of Manchester, Manchester M13 9PL, UK\\
$^{16}$ Department of Physics and Astronomy,  University of Western Cape, Cape Town, 7535, South Africa\\
$^{17}$ Department of Physics, Brown University, Providence, RI\\
$^{18}$ Radio Astronomy Lab, University of California, Berkeley, CA\\
$^{19}$ Department of Physics, University of California, Berkeley, CA\\
$^{20}$ Department of Physics and Astronomy, University of California, Los Angeles, CA\\
$^{21}$ School of Physics, University of Melbourne, Parkville, VIC 3010, Australia\\
$^{22}$ MIT Kavli Institute, Massachusetts Institute of Technology, Cambridge, MA\\
$^{23}$ Department of Physics and McGill Space Institute, McGill University, 3600 University Street, Montreal, QC H3A 2T8, Canada\\
$^{24}$ School of Chemistry and Physics, University of KwaZulu-Natal\\
$^{25}$ Scuola Normale Superiore, 56126 Pisa, PI, Italy\\
$^{26}$ Center for Astrophysics, Harvard \& Smithsonian, Cambridge, MA\\
$^{27}$ American Astronomical Society, Washington, DC
}

\date{Accepted XXX Received YYY; in original form ZZZ}

\pubyear{2022}

\begin{document}
\label{firstpage}
\pagerange{\pageref{firstpage}--\pageref{lastpage}}
\maketitle
\pagebreak
\twocolumn[
  \begin{@twocolumnfalse}
    \section*{abstract}
    \vspace{-0.2cm}
    \large
    Radio interferometers aiming to measure the power spectrum of the redshifted 21~cm line during the Epoch of Reionisation (EoR) need to achieve an unprecedented dynamic range to separate the weak signal from overwhelming foreground emissions. Calibration inaccuracies can compromise the sensitivity of these measurements to the effect that a detection of the EoR is precluded. An alternative to standard analysis techniques makes use of the closure phase, which allows one to bypass antenna-based direction-independent calibration. Similarly to standard approaches, we use a delay spectrum technique to search for the EoR signal.
    Using 94 nights of data observed with Phase~I of the Hydrogen Epoch of Reionization Array (HERA), we place approximate constraints on the 21\,cm power spectrum at $z=7.7$. We find at 95\% confidence that the 21\,cm EoR brightness temperature is $\le$(372)$^2$ "pseudo" mK$^2$ at 1.14 "pseudo" $h$\,Mpc$^{-1}$, where the "pseudo" emphasises that these limits are to be interpreted as approximations to the actual distance scales and brightness temperatures. Using a fiducial EoR model, we demonstrate the feasibility of detecting the EoR with the full array. Compared to standard methods, the closure phase processing is relatively simple, thereby providing an important independent check on results derived using visibility intensities, or related.
    
    \vspace{0.4cm}
  \end{@twocolumnfalse}
]

\begin{keywords}
techniques: interferometric -- dark ages, reionization, first stars -- intergalactic medium -- methods: statistical -- methods: data analysis.
\end{keywords}

%%%%%%%%%%%%%%%%%%%%%%%%%%%%%%%%%%%%%%%%%%%%%%%%%%

%%%%%%%%%%%%%%%%% BODY OF PAPER %%%%%%%%%%%%%%%%%%

\section{Introduction}
The Epoch of Reionisation (EoR) is a period in cosmic history during which the neutral intergalactic medium (IGM) was ionised by the first luminous sources. Current constraints on cosmic reionisation are derived from the scattering of the of the Cosmic Microwave Background (CMB) from the ionised IGM \citep{planck2018param} and from absorption effects observed in high-redshift quasars and galaxy surveys. Combined, these observations point towards a reionisation midpoint at a redshift of $z\sim7$ and a completion of reionisation by $z\sim5.5$ \citep{globalhistory}.  

The 21\,cm spin-flip emission of neutral Hydrogen (\textsc{H\,i}) will be a powerful probe of cosmic reionisation (see e.g. \cite{morales2010reionization, pritchard2012, furlanetto2016} for reviews). Once observed, the redshifted 21\,cm signal will provide spatially resolved information about the timing and duration of reionisation as well as the physical properties of the neutral IGM. In return, this will give insight into the nature of the sources providing the ionizing photons. An advantage of using the 21\,cm signal is that its cosmological redshift is an indicator of the line-of-sight distance, making it a tomographic probe of the neutral IGM \citep{tomography}. However, to increase the sensitivity to the 21\,cm signal, current interferometric experiments focus on a statistical detection by means of its power spectrum rather than tomographic imaging.  

There are several ongoing, past and future radio interferometers aimed at measuring the power spectrum of the cosmological 21\,cm signal. These include the Hydrogen Epoch of Reionization Array \citep[HERA\footnote{http://reionization.org/},][]{hera2017}, the Donald C. Backer Precision Array for Probing the Epoch of Reionization \citep[PAPER;][]{paper}, The Murchison Widefield Array \citep[MWA;][]{mwa}, the LOw Frequency ARray \citep[LOFAR;][]{lofar}, the Long Wavelength Array \citep[LWA;][]{lwa}, the Giant Metre Wave Radio Telescope \citep[GMRT;][]{gmrt} and the Square Kilometre Array \citep[SKA;][]{skaEoR}. While continuously lowering the upper limits on the 21\,cm brightness temperature of the IGM, most of these experiments are currently limited by systematic effects rather than thermal noise \citep{gmrt, cheng2018paper, kolopanis2019paper, dillon2015mwa, beardsley2016mwa, barry2019mwa, MWAII, trott2020mwa, patil2017lofar, mertens2020lofar, upperlimits, upperlimits2}. 

One particular challenge is to calibrate the instrument to the accuracy required for a detection of the cosmological 21\,cm signal. Inaccurate sky models and differences between nominally redundant baselines can introduce calibration errors that overwhelm the weak cosmological signal \citep{barry2016calibration, ewall2017calibration, byrne2019limitations, byrne2021unified}. This motivates the use of calibration-independent closure quantities \citep{TMS2017,Thyagarajan+2022,Samuel+2022} to search for the cosmological signal. In this work we use the closure phase, which is defined as the sum of the three visibility phases of an antenna triangle \citep{closurephase}. It can be shown that antenna-based direction-independent gain phases cancel in the closure phase.

The basic concept of the closure phase approach was first introduced in \cite{closure1} and its mathematical foundation is set out in \cite{closuremaths}. Using simulations, these papers confirm that the dynamic range required to detect the weak cosmological signal in the closure phase is comparable to that of a visibility based approach. Looking at HERA commissioning data, \cite{closureHERA} find that the closure phase agrees well across redundant measurements. The first results of the closure phase analysis performed on 18~nights of HERA phase~I observing are presented in \cite{closurelimits}. While the data is partially affected by systematic effects, they also identify large regions in the power spectra that are limited by thermal-like noise. 

\defcitealias{upperlimits}{H22c}
\defcitealias{upperlimits2}{H22a}

The closure phase analysis is carried out in parallel with the standard visibility based analysis \citep[cf.][hereafter \citetalias{upperlimits2}]{upperlimits2}. \citetalias{upperlimits2} report improved constraints on the 21~cm EoR power spectrum using visibility intensities from a full season of Phase~I HERA data, finding at 95\% confidence that $\Delta^2_{21} \leq (21.4)^2$\,mK$^2$ at $k=0.34\,h$\,Mpc$^{-1}$ and $z=7.9$ and $\Delta^2_{21} \leq (59.1)^2$\,mK$^2$ at $k=0.36\,h$\,Mpc$^{-1}$ and $z=10.4$. These results are an update to previously published limits obtained on a subset of the Phase~I data \citep[][hereafter \citetalias{upperlimits}]{upperlimits} and provide an improvement by factors of 2.1 and 2.6 respectively. \citetalias{upperlimits2} use a similar data processing pipeline to that of \citetalias{upperlimits}, which incorporates elaborate methods for preventing RFI and internal instrumental coupling effects from contaminating the power spectrum \citep{systematics1, systematics2}. The signal losses introduced through these non-linear processing steps are characterised in a validation pipeline that makes use of extensive simulations \citep[cf.][]{HERAvalidation}. The upper limits were then used to set constraints on the IGM and galaxies at $z\sim8$ and 10 \citep[cf.][\citetalias{upperlimits2}]{HERAtheory}. These constraints require heating above the adiabatic cooling threshold prior to $z\sim10.4$ and disfavour models with low X-ray heating. While the upper limits set by HERA are going to be increasingly important in constraining the EoR, this also calls for independent and alternative approaches to analysing the data. The closure phase analysis is one such approach which will increase the confidence in these results.

In this paper, we report upper limits of the closure phase delay power spectrum obtained from a full season of HERA Phase~I observing. The outline of this paper is as follows. In Section~\ref{sec:theory}, we shortly summarise the mathematical foundations underlying the closure phase approach. Section~\ref{sec:hera} lists the specifications of the HERA array and Section~\ref{sec:data} gives an overview of the data used in the analysis. In Section~\ref{sec:modelling}, we describe the modelling used to validate the approach. Section~\ref{sec:analysis} details the data selection and delineates the analysis pipeline. Finally, we present our results in Section~\ref{sec:results} and summarise in Section~\ref{sec:summary}.

\section{Theory}
\label{sec:theory}
The mathematical foundations of the closure phase approach are outlined in \cite{closuremaths}. For completeness, we summarise some of the mathematical formalism most relevant to the closure phase delay power spectrum  analysis, without repeating the involved mathematics in \cite{closuremaths}. 

In the limit where the cosmological 21\,cm signal is weak relative to the foreground continuum visibility amplitudes, we can treat the cosmological signal as a small perturbation to the foreground visibility phase $\phi_p$ of baseline $p$. In a first-order approximation, it can be shown that these phase perturbations are \citep{closuremaths}

\begin{equation}
\label{eq:dphi}
    \delta \phi_p(\nu) \approx \Im\left\{\frac{V^\mathrm{P}_p(\nu)}{V^\mathrm{F}_p(\nu)}\right\},
\end{equation}
where $V^\mathrm{P}_p$ and $V^\mathrm{F}_p$ are the perturbing visibility and the foreground visibility of baseline $p$ respectively and $\Im$ denotes the imaginary part. The closure phase $\phi$ is the sum of the three visibilitiy phases of a closed antenna triangle (triad) and its perturbation is simply the sum of the corresponding phase perturbations,

\begin{equation}
\label{eq:dphi}
    \delta \phi(\nu) \approx \sum_{p=1}^3 \Im\left\{\frac{V^\mathrm{P}_p(\nu)}{V^\mathrm{F}_p(\nu)}\right\}.
    %=\frac{1}{2}\sum_{p=1}^3 \left(\frac{V^\mathrm{P}_p(\nu)}{V^\mathrm{F}_p(\nu)} - \frac{\overline{V^\mathrm{P}_p(\nu)}}{\overline{V^\mathrm{F}_p(\nu)}}\right)
\end{equation}

The line-of-sight fluctuations of the cosmological signal will cause the phase perturbations to fluctuate across frequency, while the foreground visibility phases are due to broad-band continuum emission, and relatively smooth in frequency. It is this frequency dependence that allows us to separate the cosmological signal from the bright continuum foregrounds. More precisely, we perform a Fourier transform of the closure phase along frequency, which is known as a delay transform \citep[cf.][]{dspec}. In the delay spectrum the spectrally smooth foregrounds are confined to low delay modes, while higher delay modes can be used to set upper limits on the \textsc{H\,i} 21\,cm emission from the EoR. The latter is known as the EoR window \citep{eorwindowmath,datta}. Formally, we define the delay spectrum as

\begin{equation}
\label{eq:dspec}
    \widetilde{\Psi}_\triangledown(\tau) = V_\mathrm{eff}\int e^{i\phi(\nu)}W(\nu)e^{i2\pi \nu\tau}\mathrm{d}\nu,
\end{equation}
where $V_\mathrm{eff}$ is a scaling factor, $W$ is a spectral tapering function shaped to fit the observed band, $\nu$ is the frequency and $\tau$ is the delay. In this analysis, we use a Blackman-Harris function \citep{blackman-harris} for $W$, which is suited to the high  dynamic range requirements of the measurement. Note that instead of directly transforming the closure phase $\phi$, we transform its complex exponential. Doing this, we avoid the discontinuities that can arise because of the circularity of phase. 

The effective visibility $V_\mathrm{eff}$ provides the delay spectrum with units of Jy\,Hz, which are the units of a standard visibility delay spectrum. Furthermore, $V_\mathrm{eff}$ should be designed to gauge the strength of the closure phase fluctuations to the strength of the perturbing signal. That is, it should counteract the inverse proportionality to the foreground visibility in equation~\ref{eq:dphi} so that the spectral fluctuations in $\delta\phi$ are of similar strength to the fluctuations in $V_p^\mathrm{P}$. This allows one to combine measurements of regions on the sky with different foreground visibility amplitudes. Motivated by this, we define

\begin{equation}
    V_\mathrm{eff}^{-2} = \sum_{p=1}^3 {\left(\widehat{V}_p^\mathrm{F}\right)}^{-2},
\end{equation}
where $\widehat{V}_p^\mathrm{F}$ is an estimate of the foreground visibility amplitude weighted by the window function $W$ and averaged over the observed subband \citep[cf.][]{closuremaths}. The summation in inverse quadrature gives weight to the baselines with the smallest foreground visibility amplitudes, where the spectral fluctuations of the 21\,cm signal are expected to be strongest. In this analysis, we use calibrated data to estimate $V_p^\mathrm{F}$, as it is readily available from the visibility processing pipeline. However, in principle, we could use an accurate sky model instead, making this approach completely independent of calibration. 
Note that $V_\mathrm{eff}$ is deliberately chosen to be frequency independent to avoid further foreground contamination into the EoR window. This assumption should be reasonable over the relative narrow bands considered herein \citep{closuremaths}. Regardless, any claimed limits or detection using this technique comes down to comparison with physical models for the cosmological and foreground signals, which are treated identically to the real data.

Ultimately, we are interested in the power spectrum $|\widetilde{\Psi}_\triangledown(\tau)|^2$, which, statistically, is independent of direction and polarisation. This is due to the assumed isotropy and polarisation-independence of the 21\,cm signal and allows for the incoherent averaging of different portions of the data. Section~\ref{sec:avg2} describes how we do this in practice. 

\begin{figure}
    \centering
    \includegraphics[width=\linewidth]{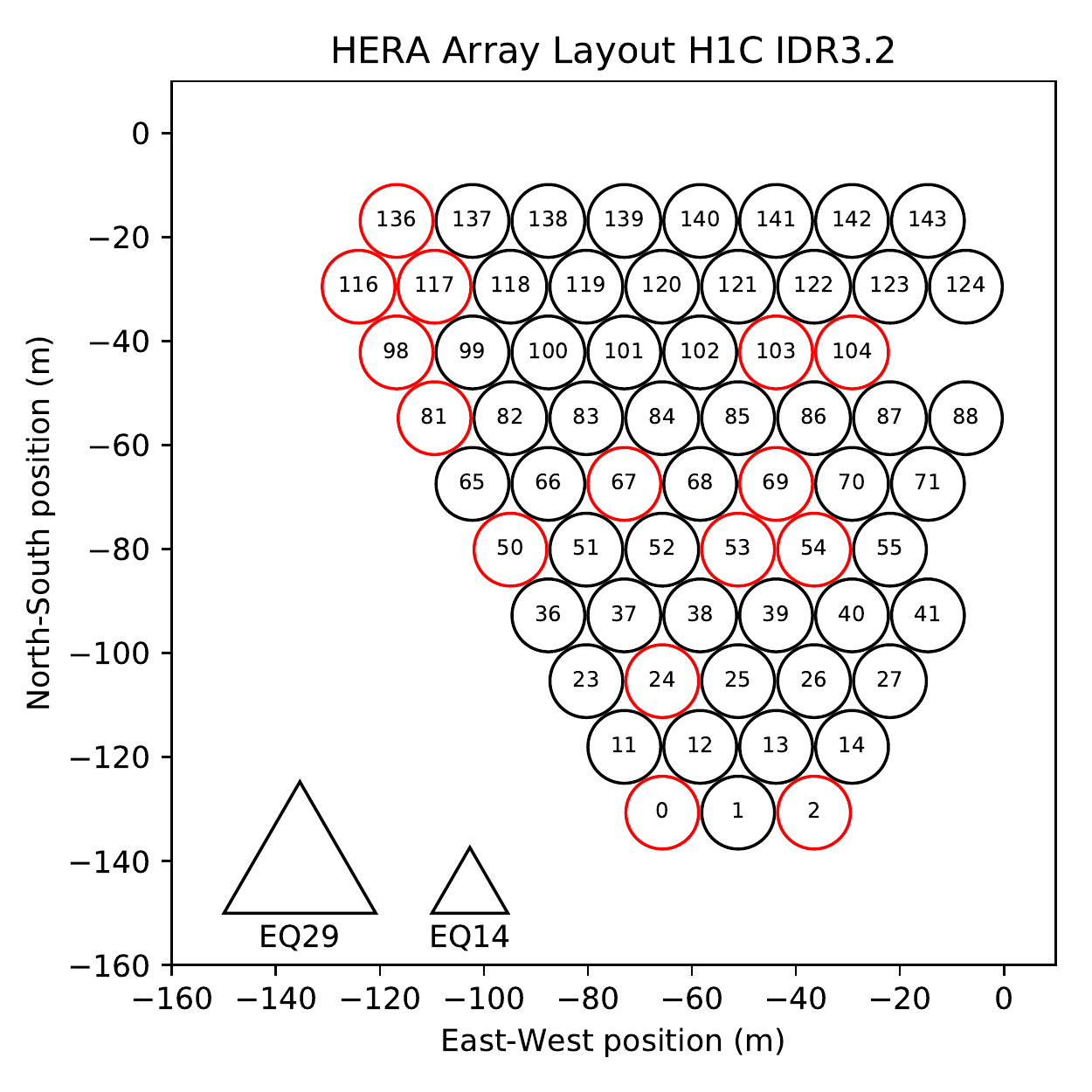}
    \caption{The layout of the antennas used in this analysis. These antennas form a part of the south-west section of the hexagonal array. Their diameter is 14.0 metres and the separation between two neighbours is 14.6 metres. Antennas that were flagged completely using flags from \citetalias{upperlimits2} are shown in red (cf. section \ref{sec:flags}). The equilateral triad shapes (EQ14 and EQ29) used in this analysis are shown in the bottom left corner of this plot.}
    \label{fig:array}
\end{figure}

\section{HERA}
\label{sec:hera}
HERA is a low-frequency radio interferometer designed to measure the 21\,cm emissions of neutral Hydrogen during the EoR. Located in a radio quiet zone in the Karoo desert in South Africa, it is minimally affected by radio frequency interference (RFI). In its complete form, HERA will consist of 320 closely packed 14-metre dishes arranged in a split-hexagonal core and complemented with 30 outrigger antennas \citep{hera2017}. The highly redundant layout allows for high-precision redundant calibration and is optimized for the delay spectrum approach \citep{redundantConfig}. Furthermore, it is ideally suited to the closure phase analysis presented herein, due to the many redundant closure triads in the array. 

The data used in this analysis was taken with the HERA Phase~I system, which re-used several system elements from its precursor instrument PAPER. These elements include dipole feeds, parts of the analog signal chain and the correlator. The Phase~I system operated at frequencies between 100 and 200\,MHz and observed 1024 channels simultaneously, resulting in a spectral resolution of 97.7\,kHz. The temporal resolution is 10.7\,s \citep[][\citetalias{upperlimits}]{hera2017}.

\section{Data}
\begin{figure*}
    \centering
    \includegraphics[width=0.9\linewidth]{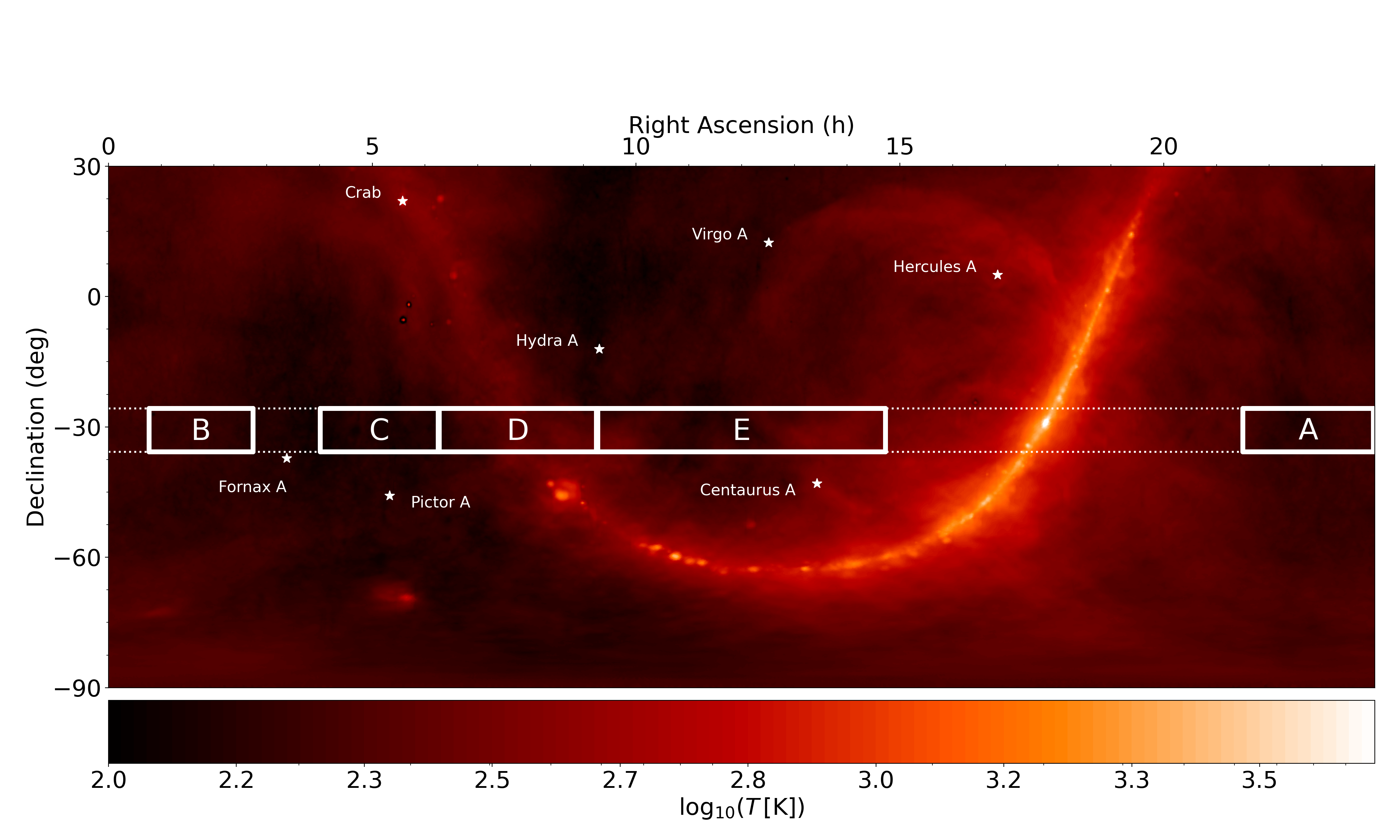}
    \caption{The observed fields A, B, C, D and E plotted on the global sky model (GSM) at 160\,MHz from \protect\cite{gsm16}. The locations of other bright radio sources are also shown on the map. The dashed lines indicate the $\sim$10$^\circ$ field-of-view of HERA at 150\,MHz. The fields are chosen to avoid the Galactic plane and the bright radio galaxy Fornax~A, which is located only $\sim 6^\circ$ from the centre of the HERA strip.}
    \label{fig:map}
\end{figure*}
\label{sec:data}
The observations used for this paper cover 94 nights between the Julian Dates 2458041 and 2458208 (15$^\mathrm{th}$ October 2017 to 1$^\mathrm{st}$ April 2018), during which the array was under active construction. As a result, the number of antennas changes throughout the observing season. All together, we use 48 antennas, which constitute a part of the south-west segment of the final array (see Figure~\ref{fig:array}). These antennas are flagged on a nightly basis using the flags from \citetalias{upperlimits2} resulting in 35 to 41 unflagged antennas at any one night. The per antenna and per night flags are further detailed in Section~\ref{sec:flags}.

In this paper we only consider measurements on closed antenna triads, rather than baselines or single antennas. We employ closed triad classes that are equal or point-symmetric at the origin in $uv$-space. For example, a north facing equilateral triad and a south facing equilateral triad belong to the same class, as their $uv$-geometry can be matched by inversion through the origin (i.e. they are conjugates of one another). As a result their closure phases differ only by sign. We perform the analysis on two triad classes individually, equilateral 14.6-metre and 29.2-metre triads, which we refer to as EQ14 and EQ29 respectively. This will allow us to see if the two triad classes are affected differently by systematic effects and to what extent their different responses to the foregrounds influence the final power spectra. Ultimately, they probe different spatial scales, leading to more stringent constraints on cosmic reionisation.

\begin{table}
\renewcommand{\arraystretch}{1.2} 
\caption{The LST ranges of the fields used in this work and their total observation time in hours.}
\begin{tabular}{l|ccccc}
\hline\hline
Field     & A          & B           & C          & D          & E \\  
LST (h) & 21.5-0.0 & 0.75-2.75 & 4.0-6.25 & 6.25-9.25 & 9.25-14.75 \\
Total (h) & 55 & 89 & 148 & 214 & 210\\
\hline\hline
\end{tabular}
\label{tab:fields}
\end{table}

As a zenith pointing array, the observable portion of the sky is limited to a strip centred at a declination of -30.7$^\circ$. The width of this strip is defined by the Full Width at Half Maximum (FWHM) of an antenna beam. For a HERA dish, the FWHM is approximately 10$^\circ$ at 150\,MHz \citep{beam}. As shown in Figure~\ref{fig:map} the Galactic centre transits overhead at Local Sidereal Time (LST) $\sim$18\,h. Moreover, the bright radio galaxy Fornax~A at $\sim$3.3\,h RA and -37$^\circ$ DEC is located close enough to the HERA strip to produce a sizable fraction of the total power received by an antenna. We chose to analyse observations from five fields, denoted A, B, C, D and E, that avoid these bright regions on the sky and hence limit the dynamic range required to measure the cosmological 21\,cm signal. The LST-ranges of the observed fields and their total observation time are listed in Table~\ref{tab:fields}.

In this paper, we report results for a frequency band ranging from 160.59 to 167.97\,MHz corresponding to a central redshift of about 7.7. This band overlaps with Band~2 (152.25–167.97\,MHz) used in the visibility processing \citepalias{upperlimits2}. The reason for the trimming of Band~2 is that we found residual RFI in the lower part of the band after averaging the closure phases. The inclusion of such RFI can lead to excess power in the power spectrum, making it indistinguishable from a sky based signal. The evidence by which we decided to cut the band is further discussed in Appendix~\ref{sec:rfi}. Note that the initial bandwidth of 7.4\,MHz of our band is reduced to an effective bandwidth of 3.7\,MHz after applying the Blackman-Harris function in equation~\ref{eq:dspec}.

\section{Modelling}
\label{sec:modelling}
We use data simulations to validate the closure phase approach and ultimately to compare the data and the expected signal of a given EoR model. As our measurements are currently limited by noise and systematics, we use this comparison to estimate the additional sensitivity needed to achieve a detection of a fiducial EoR model \citep{eos}. In the following two sections we describe the sky models used here, consisting of foregrounds, a 21\,cm signal component and noise.

\subsection{Foregrounds}
\label{sec:foregrounds}
We use the GaLactic and Extragalactic All-Sky MWA survey catalogue \citep[GLEAM,][]{gleam} as a basis for our foreground models. Since the GLEAM catalogue does not cover the galactic plane, we restrict our simulations to fields A, B and C of this analysis. For a given LST, we select GLEAM point sources that have an integrated flux density exceeding 50\,mJy at 151\,MHz and fall within a radius of 15$^\circ$ of the pointing centre (i.e. they lie within the main lobe of the antenna beam). Flux densities are then interpolated in frequency using the fitted spectral indices and fluxes provided in the catalogue. Where no spectral index is provided, we fit it using the integrated flux densities at 122, 130, 143, 151, 158, 166, and 174~MHz, respectively, assuming that the flux is described by a power law in that frequency range. Applying a discrete version of the van Cittert-Zernike theorem \citep{cittert1934wahrscheinliche, zernike1938concept}, we compute the visibilities $V_p$ for two antennas separated by $\textbf{b}_p$ as

\begin{equation}
    V_p = \sum_i A\left(\textbf{s}_i, \nu\right) I_i\left(\nu\right) 
    e^{2\pi i \nu \frac{\textbf{b}_p \cdot \textbf{s}_i}{c}},
\end{equation}
where $c$ is the speed of light, $I_i\left(\nu\right)$ is the flux density of source $i$ at a location given by the unit vector $\textbf{s}_i$ and at frequency $\nu$ and $A\left(\textbf{s}_i, \nu\right)$ is the response of a simulated HERA-beam \citep{beam}.

Several strong radio sources have been peeled from the GLEAM catalogue \citep[see Table 2 in][]{gleam}. We simulate visibilities for these sources whenever they are above the horizon and assume that they are point-like, which is justified for the short baselines used in this analysis. For Fornax~A we use a three component model informed by \cite{FornaxA}. The model consists of point sources representing the two lobes and the core with flux densities of 478 Jy (west), 260 Jy (east) and 12 Jy respectively at 154\,MHz. The spectral indices of the lobes are -0.77 while that of the core is -0.88. For the other peeled sources we use the spectral indices and total flux densities provided in Table 2 of \cite{gleam}. 

\subsection{EoR signal}
We use the public "Faint Galaxies" simulation output of 21cmFast as an EoR model \citep{eos} and follow a similar procedure to the one used in \cite{closurelimits}. The EoR light cone consists of two transverse axes and one line-of-sight axis which each have a length of 1.6\,~cMpc and a resolution of 1024 voxels. In our simplified model we identify the transverse direction with orthographically projected angular extent and the line-of-sight direction with frequency. At $z\sim~7.8$ the corresponding angular resolution is thus $\sim$28$\mathrm{'}$ and the frequency resolution $\sim$93\,~kHz, where the conversion was done using the cosmological parameter from \cite{planck2018param}. Subtending an angle of $\sim$10$^\circ$, the EoR light cube does not cover the entire HERA field of view. For this reason, we tile the model three times along the transverse axes such that each tile is a reflection of its neighbouring tiles, thus avoiding discontinuities at the tiling boundaries. To reduce subsequent computing time, we then smooth and downsample the model to an effective angular resolution of $\sim$7$\mathrm{'}$ corresponding to 256 pixels per side. The resolution thus obtained is still well below the FWHM of $\sim$5$^\circ$ of the synthesised HERA beam. Treating each pixel as a point source, we compute visibilities in the same way as for the foregrounds described in Section~\ref{sec:foregrounds}.

\subsection{Noise}
\begin{figure}
    \centering
    \includegraphics[width=\linewidth]{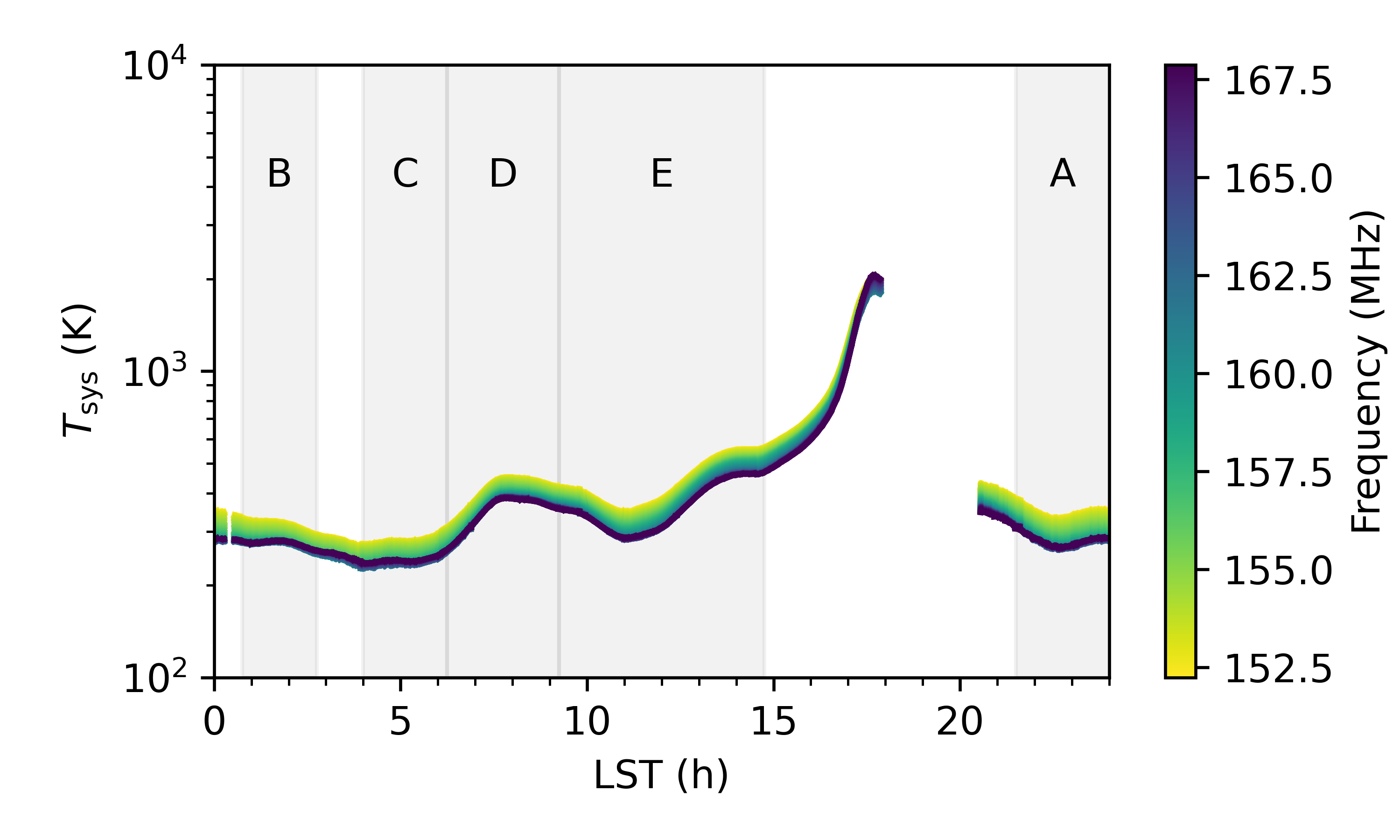}
    \caption{Our model of the system temperature $T_\mathrm{sys}$ as a function of LST and frequency. The observed fields are shaded in grey.}
    \label{fig:Tsys}
\end{figure}
\begin{figure*}
    \centering
    \includegraphics[width=0.9\linewidth]{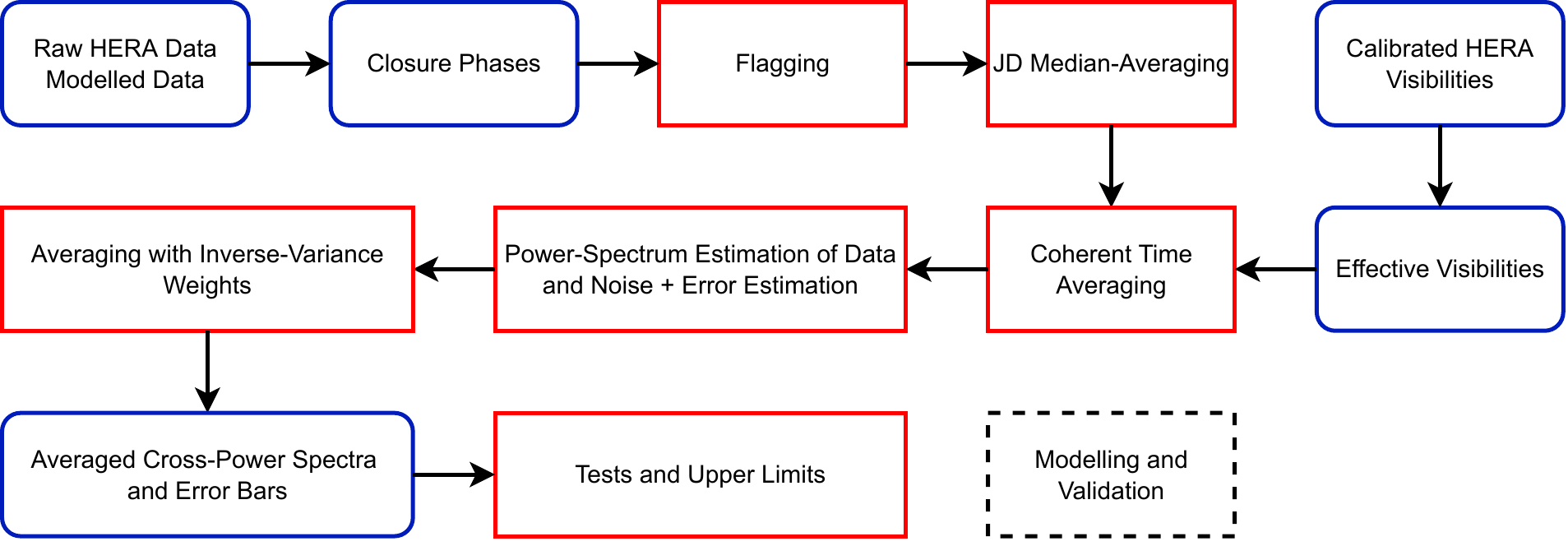}
    \caption{A flowchart showing the processing steps (red) and data products (blue) of the closure phase analysis. The closure phases are computed directly from raw HERA data, i.e. there is no calibration step in the pipeline. The modelling and validation steps (black dashed) are described in Section~\ref{sec:modelling} and Appendix~\ref{sec:lstavg} and are performed independently from the data processing steps.}
    \label{fig:flowchart}
\end{figure*}
We use calibrated auto-correlation visibilities, $V_{p}^\mathrm{auto}$, to model the system temperature of the array, $T_\mathrm{sys}$, as a function of time $t$ and frequency $\nu$ \citep[cf.][]{errorbar}. The system temperature of a single antenna, indexed by $p$, is calculated as

\begin{equation}
    T_{\mathrm{sys}, p}(t, \nu) = \frac{c^2 V_{p}^\mathrm{auto}(t, \nu)}{2k_b\nu^2\Omega},
\end{equation}
where $\Omega$ is the integrated beam area and $k_b$ is the Boltzmann constant and $V_p$ is taken from LST-binned and systematics filtered auto-correlations of the visibility processing pipeline \citepalias{upperlimits2}. The values of $T_{\mathrm{sys}, p}$ vary strongly across antennas, with an average relative standard deviation of 17\%. To obtain a model of the system temperature that characterises the whole array, we calculate the quadratic average of $T_{\mathrm{sys}, p}$ over all antennas. Figure~\ref{fig:Tsys} shows the model of $T_\mathrm{sys}$ plotted against LST and coloured by frequency. As expected, $T_\mathrm{sys}$ rises around the transit of the galactic plane and is at its lowest in the colder fields, A, B and C.

The standard deviation of the visibility noise is related to $T_\mathrm{sys}$ through the radiometer equation

\begin{equation}
    \sigma(t, \nu) = \frac{\sqrt{2} k_b \nu^2 \Omega T_{\mathrm{sys}}(t, \nu)}{c^2 \sqrt{\Delta t \Delta \nu}},
\end{equation}
where $\Delta t$ and $\Delta \nu$ are the integration time and frequency resolution respectively. In our model, we draw the noise from a Gaussian distribution with standard deviation $\sigma$ and add it to the real and imaginary parts of the simulated visibilities individually. These visibilities are then propagated through the data analysis pipeline described below (Section~\ref{sec:analysis}) using the same flags as for the data.

\section{Analysis}
\label{sec:analysis}
This section provides an overview of our analysis pipeline. We start by explaining how we obtain closure phases from raw HERA data and how we apply flags to the data. We then continue with our averaging techniques as well as the power spectrum and error estimation. Lastly, we describe how we estimate upper limits. The processing steps and the data products of this analysis are delineated in Figure~\ref{fig:flowchart}.

\subsection{Computing Closure Phases}
Since closure phases are independent from antenna-based calibration, we compute them directly from raw visibility data. For a given triad, we do this by taking the phase of the triple product of the three visibilities. We then generate a uniformly spaced LST-grid with a resolution given by the integration time of a measurement and replace the time stamps of the closure phases with their nearest points on the grid. This allows us to combine the closure phases to a five dimensional data array of shape $N_p \times N_j \times N_t \times N_l \times N_f$. In this analysis the number of polarisations is $N_p=2$, as we only keep the parallel "East-West" and "North-South" polarisation products and the number of frequencies is $N_f=76$ for our chosen band. The number of nights $N_j$, triads $N_t$ and LST-integrations $N_l$ depend on the observed field (see Section~\ref{sec:data}). Note that the closure phases will later be averaged across nights, so the effect of LST-gridding on the power spectrum will be similar to that of coherent time averaging described in Appendix \ref{sec:lstavg}. The introduced signal loss is expected to be negligible, since the gridding interval is equal to the time interval of a single integration.

\subsection{Data Flagging}
\label{sec:flags}
We take the nightly antenna flags from \citetalias{upperlimits2} to flag triads formed by such antennas. These flags are informed by specially designed metrics for detecting malfunctioning antennas \citep{antmetrics} and by the redundant-baseline calibration process of \citetalias{upperlimits2} which is able to identify particularly non-redundant antennas \citep[cf.][]{redcal}. The exact reason for flagging individual antennas can be found in \cite{memo97} and the notebooks referenced therein. We further adopt the time flags listed in Table 2 and 3 in \citetalias{upperlimits2} that are due to broadband RFI and digital system failures. Time flags that are due to calibration issues are not applied, as these may not be relevant to the closure phase analysis. We also flag times during which the sun is above the horizon and LST's at which $V_\mathrm{eff}<5$\,Jy. Setting this condition on $V_\mathrm{eff}$, we avoid instances where one of the visibility amplitudes happens to be close to zero or is of the order of the thermal noise because of the coincidence of the orientation of foreground sources. Firstly, the closure phase is poorly defined if one of the visibility amplitudes is zero. We observed that this can cause poles in the closure phase spectrogram, which in return can lead to excess power in the power spectrum. Secondly, the noise variance of the scaled closure phase will tend to zero together with $V_\mathrm{eff}$. This has a strong effect further down the analysis pipeline (see Section~\ref{sec:avg2}), where we compute an inverse variance weighted average across LST. That is, LST's at which $V_\mathrm{eff}$ is small relative to the visibility noise variance will be weighted disproportionately high relative to other LST's. We find that a threshold of $V_\mathrm{eff} \sim 5$\,Jy eliminates these unwanted effects. Across all times, both polarisation products and both triad classes the data is flagged 7\% of the time because of this condition. For model data, we find this flagging condition to be inadequate, since the visibility amplitudes are generally lower due to unmodelled diffuse emission. We therefore flag the model data by hand in regions where one of the visibility amplitudes crosses zero, resulting in a flagging frequency of 6\%. We use the model flags on the data and vice versa to prevent a bias between the two.

Other than that, we do not flag data on a per-integration and per-frequency basis and rely instead on robust averaging to filter out remaining RFI (see Section~\ref{sec:avg1}). In particular, we do not use the final "by-hand" flags of \citetalias{upperlimits2}. 

\begin{figure*}
    \centering
    \includegraphics[width=0.48\linewidth]{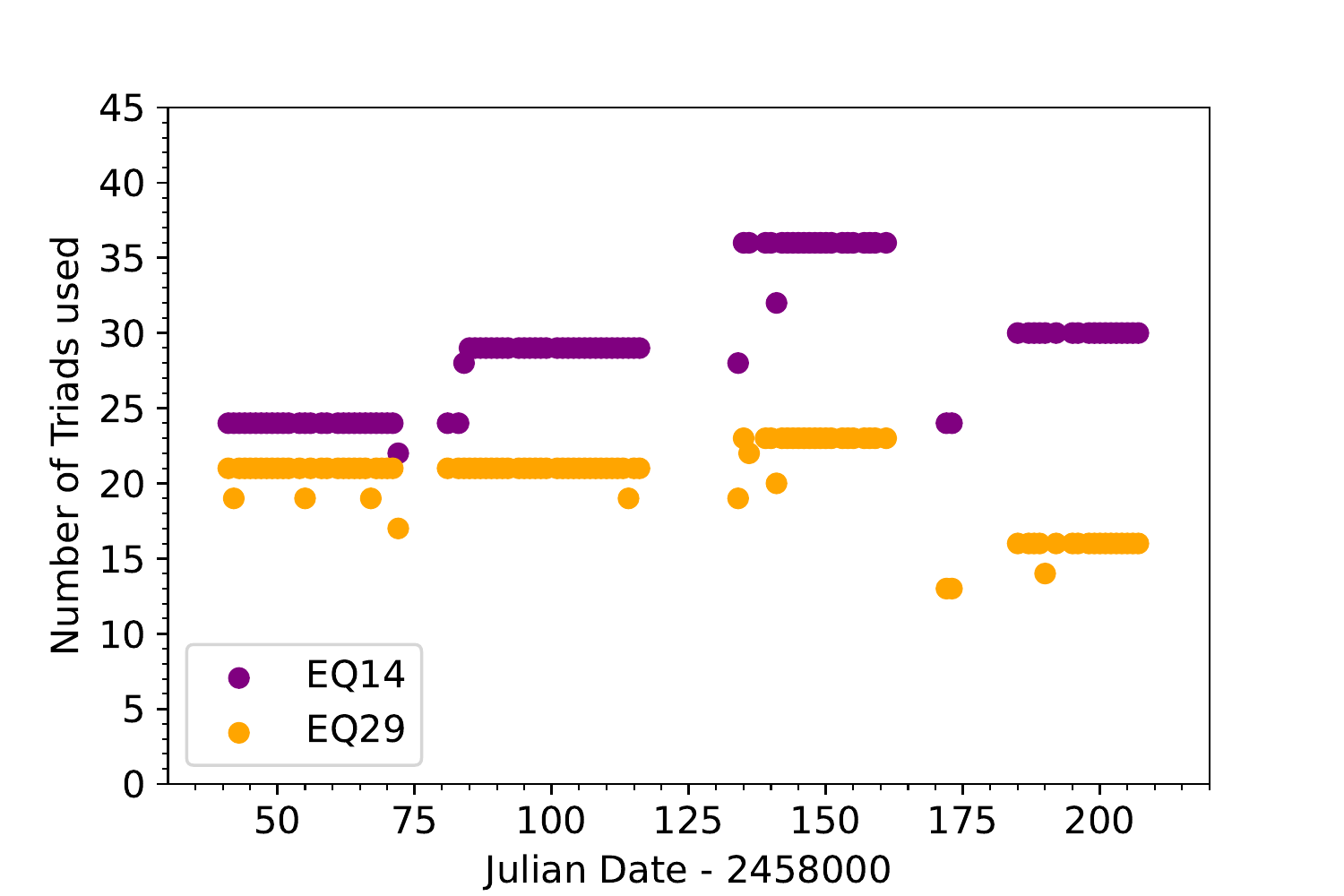}
    \includegraphics[width=0.48\linewidth]{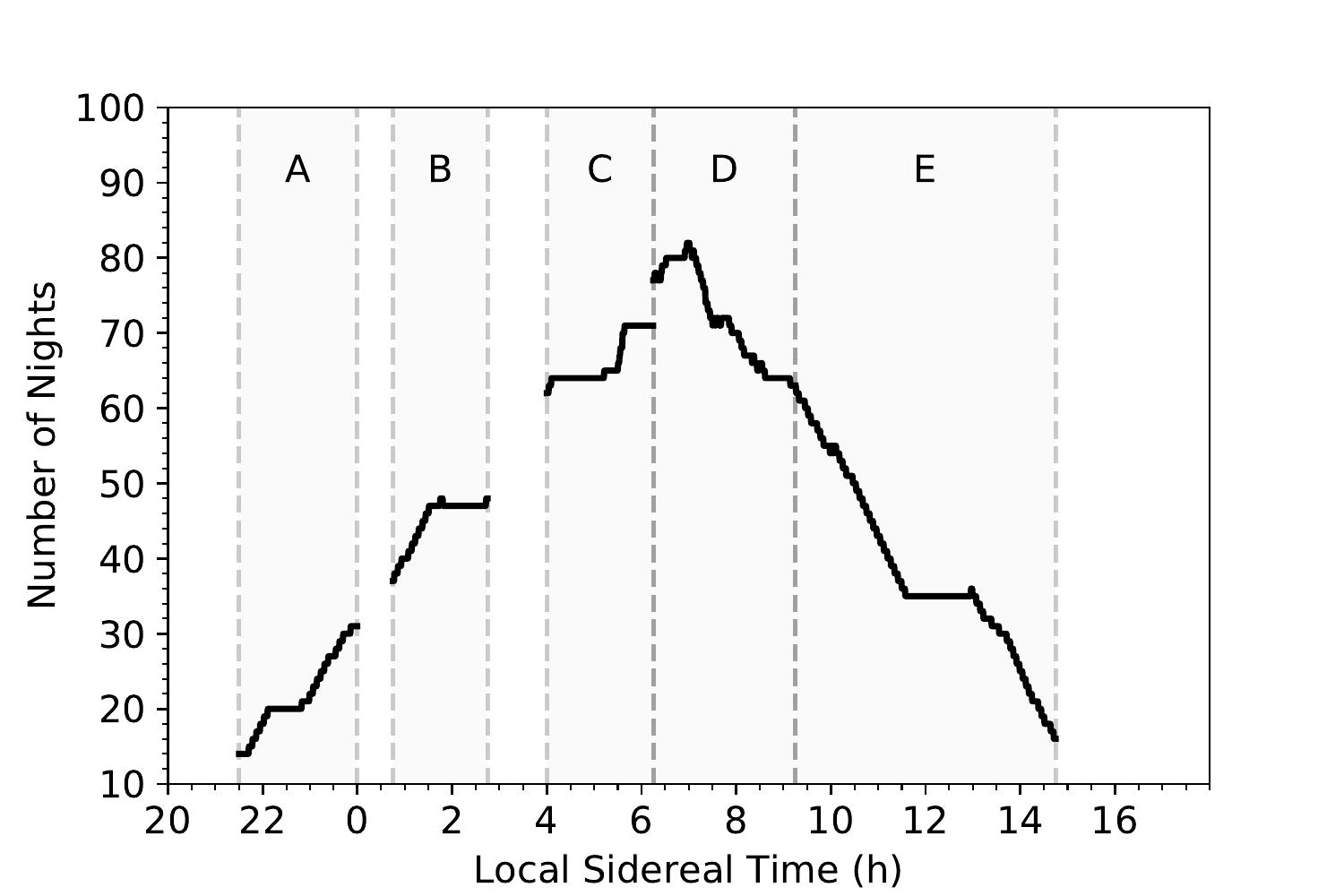}
    \caption{The number of triads after flagging for a given triad class and parallel polarisation product (East-West, North-South) as a function of Julian Date (left) and the number of nights covering a given Local Sidereal Time for fields A, B, C, D and E (right). In the first three epochs, the number of triads increases as more antennas were put into operation. }
    \label{fig:obs}
\end{figure*}

Figure~\ref{fig:obs} shows the amount of data used after flagging. As can be seen in the left plot, the observing season can be divided into four epochs, where, except for the fourth epoch, each successive epoch has an increased number of triads in use. The reason for the decreased number of triads in the fourth epoch can be traced to an increased number of malfunctioning antennas, despite there being more antennas connected \citep[cf.][and notebooks referenced therein]{memo97}. The number of triads ranges from 13 (EQ29, Epoch 4) to 36 (EQ14, Epoch 4). Note that some of these triads share a baseline and hence do not have independent noise. The right plot in Figure~\ref{fig:obs} shows the number of nights covering a given LST, ranging from 14 night at $\sim$22\,h LST to 82 unflagged nights at $\sim$ 7\,h LST.

\subsection{Averaging Closure Phases}
\label{sec:avg1}
\begin{figure}
    \centering
    \includegraphics[width=\linewidth]{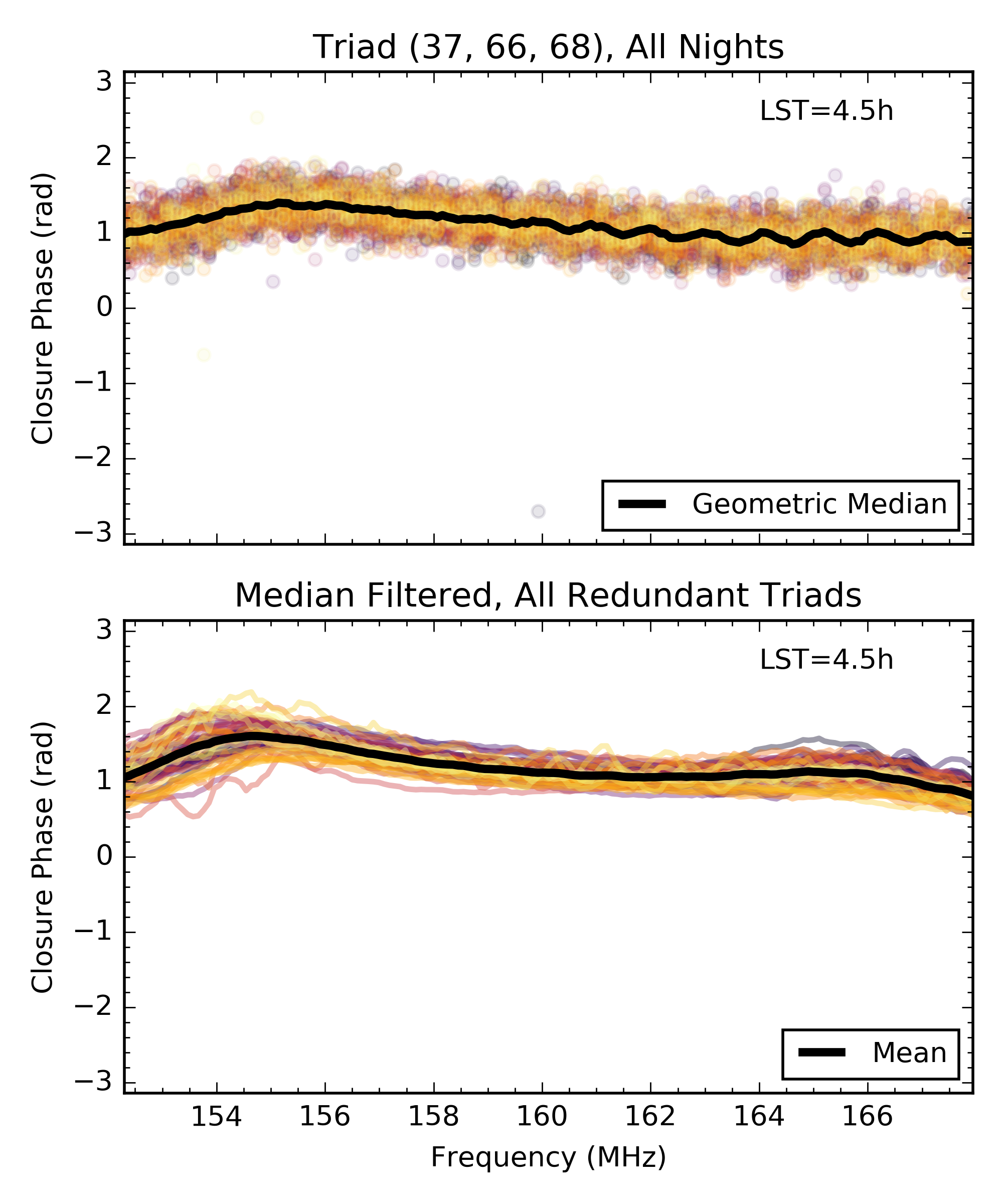}
    \caption{Closure phase spectra of data used in this analysis. The top plot shows the spectra for one triad across all nights and the median-averaged and time-averaged spectrum in black. The night-to-night variations can be attributed to thermal noise and RFI outliers. The averaged spectrum reveals a sinusoidal systematic effect with a period of $\sim$1\,MHz. The bottom plot shows the filtered and averaged spectra for all triads. Averaging redundant triads reduces the spectral ripple considerably (black line).}
    \label{fig:cpspec}
\end{figure}

Data containing the same sky-based signal can be averaged coherently. Under the assumption of perfect redundancy, the closure phase data has two axes suited for coherent averaging, namely, the repeated LST-integration of different nights and triads from the same class. Indeed, the night-to-night variations of the closure phase can be attributed to noise and RFI. The variations among nominally redundant triads, on the other hand, are clearly non-random and are caused by non-redundancies (see Figure~\ref{fig:cpspec}). For visibilities, the loss of sensitivity to the cosmological signal due to non-redundancy has been found to be at a level of 1-2 percent \citep[][\citetalias{upperlimits2}]{redundancy2}. It has yet to be established, if these estimates also apply to the closure phase, but, for the time being, we assume that the loss of sensitivity under averaging of nominally redundant triads is at a similar level. 

In our processing pipeline, we first average the complex exponentials of the closure phases across nights. This is done using a geometric median, which is defined as the data point that minimises its Euclidean distance to all other data points on the complex plane, i.e.

\begin{equation}
    \widehat{\phi} = \underset{\phi_n}{\mathrm{arg\,min}}\sum_{m}{\left|e^{i\phi_n} - e^{i\phi_{m}}\right|},
\end{equation}
where the subscripts stand for the Julian Dates.
This estimator has the advantage of being robust to outliers, such as RFI, while also respecting the circularity of the phase. The application of the geometric median is the only measure we take to mitigate the imprint of RFI on the final power spectrum. A persistent or repeated signal will not be eliminated by this treatment, therefore necessitating a band that is free of such effects (cf. Section~\ref{sec:data} and Appendix~\ref{sec:rfi}).

Note that in the visibility processing \citepalias{upperlimits2} the four epochs (see Section~\ref{sec:flags}) are at first averaged independently to allow for better systematics mitigation and statistical tests. As these processing steps are not part of this analysis, we include all epochs in the median-average. This increases the efficiency of the median at rejecting outliers.

Next in the pipeline, we average some neighbouring LST-integrations. Here, we average in intervals of 171.2\,s (16 integrations) using the arithmetic mean. The averaging in time is justified by the invariance of the closure phase to the translation of the sky. However, the antenna beam breaks this symmetry, as a result of which we expect the averaging to introduce a loss of sensitivity to the cosmological signal. Using our model data, we established the expected scale of this loss to be $\sim$2\%. The method we used to determine this loss is detailed in Appendix~\ref{sec:lstavg} of this paper.

At this point, we could, in principle, average the closure phases of redundant triads. However, deferring this averaging until after the computation of cross-power spectra will allow us to omit cross-terms between triads that share a baseline (see Section~\ref{sec:avg2}). This has the advantage of mitigating the effect of baseline based systematics that are coherent across nights \citep[see][]{closurelimits}.

\subsection{Forming Cross-Power Spectra}
\label{sec:xps}
We divide the data into two bins containing nights with odd and even Julian Dates respectively and perform the averaging described in Section~\ref{sec:avg1} separately in each bin. This allows us to compute cross-power spectra between the two bins, defined as 
\begin{equation}
    P_\triangledown(\kappa_\parallel) = \left( \frac{c^2}{2 k_B \nu^2} \right)^2
    \left( \frac{X^2Y}{\Omega B_\mathrm{eff}}\right) \times \widetilde{\Psi}_\triangledown(\tau) \overline{\widetilde{\Psi}_\triangledown^{'}}(\tau),
\end{equation}
where $\widetilde{\Psi}_\triangledown$ and $\widetilde{\Psi}_\triangledown^{'}$ are two delay spectra drawn from the first and second bin respectively and the bar denotes complex conjugation. The scaling factor is taken from the standard visibility power spectrum, where the first bracket converts flux density to brightness temperature, $X^2Y$ converts spectral and angular units to cosmological distances \citep[e.g.][]{eorwindowmath}, and $\Omega$ and $B_\mathrm{eff}$ normalise the power by the integrated squared beam response and the effective bandwidth respectively. With this definition, the cross-power spectrum has units of "pseudo" mK$^2$\,(Mpc/h)$^{-3}$ and is a function of $\kappa_\parallel = 2\pi \tau / X$, which has units of "pseudo" (Mpc/h)$^{-1}$. In line with the practice introduced in \cite{closurelimits}, we use the "pseudo" to emphasise that these units are not physical but are instead used as approximations to the real distance scales and brightness temperatures \citep[see][]{closuremaths}. Note that in computing the scaling factor $X^2Y$ we require cosmological parameters. We take these parameters from \cite{planck2018param} and use $H_0=100\,h$ (km/s)/Mpc for the Hubble constant.

Ideally, the two delay spectra $\widetilde{\Psi}_\triangledown$ and $\widetilde{\Psi}_\triangledown^{'}$ should contain the same contribution due to sky, but independent thermal noise and transient systematic effects. The sky should therefore occupy the positive real part of $P_\triangledown(\kappa_\parallel)$, while the noise can introduce negative and imaginary components. Hence, we can further reduce the noise by averaging different cross-power spectra for which, statistically, the cosmological signal has the same underlying power spectrum. This averaging is further detailed in the next section. 

In Section~\ref{sec:results}, we also present a form of the power spectrum which was first introduced in \cite{closurelimits} and is mathematically akin to the widely used cosmological variance. It is computed as $\Delta_\triangledown^2(\kappa) = \kappa^3 P_\triangledown(\kappa_{||}) / (2\pi^2)$, where the units are mK$^2$ and $\kappa^2 = \kappa_{||}^2 + \kappa_{\perp}^2$. The perpendicular component is defined in analogy to the perpendicular $k$-modes in the visibility analysis as $\kappa_{\perp}=2\pi(|\mathbf{b}|/\lambda)/Y$, where $|\mathbf{b}|$ is the baseline length of an equilateral triad and $\lambda$ is the wavelength. In the context of the closure phase analysis, this definition of $\kappa$ is only adequate for equilateral triad classes, where all baselines are sensitive to the same perpendicular $k$-modes \citep[cf.][]{closuremaths}. However, in this paper it will be useful for obtaining upper limits that correspond approximately to the cosmological variance and, thus, can be compared to the results of standard analysis techniques.

\subsection{Averaging Cross-Power Spectra}
\label{sec:avg2}
After forming cross-power spectra the data for a given polarisation product has shape $N_t \times N_t \times N_l \times N_\tau$ where $N_\tau=N_f$ is the number of delays. We precede with averaging across the two redundant triad axes and the LST axis, noting that the later is an incoherent average making use of the assumed isotropy of the 21\,cm signal. 

The noise as well as systematic effects vary across triads and by the pointing direction of the telescope, which for a zenith pointing array, such as HERA, coincides with the LST. This motivates a weighted average 
\begin{equation}
    \widetilde{\Psi}_\triangledown(\tau) \overline{\widetilde{\Psi}_\triangledown^{'}}(\tau) = \sum_{i,j,t} w_{ijt} \widetilde{\Psi}_\triangledown(i,t,\tau) \overline{\widetilde{\Psi}_\triangledown^{'}}(j,t,\tau),
\end{equation}
where we choose the normalised weights $w_{ijt}$ so that
\begin{equation}
    w_{ijt} \propto 
    \begin{cases}
    w_{it}w_{jt}, & i\mathrm{\ and\ }j\mathrm{\ do\ not\ share\ a\ baseline.} \\
    0, & i\mathrm{\ and\ }j\mathrm{\ share\ a\ baseline.}
    \end{cases},
\end{equation}
where $w_{it}$ is the inverse variance of triad $i$ at time $t$. We estimate $w_{it}$ by differencing the scaled complex closure phases from different nightly bins and computing the variance along the frequency axis. Omitting cross-terms between triads with shared baselines prevents baseline based systematic that are coherent across nights from contaminating the power spectrum. This effect has been observed in \cite{closurelimits}, where the cross-power between identical triads produced a positive bias in the high-delay region of the power spectrum. 

Assuming an unpolarised and isotropic 21\,cm signal, we further average over the two polarisation products and in bins of $|\kappa_{||}|$. The cross-power spectra thus obtained are the ones presented in Section~\ref{sec:results}.

\subsection{Estimating Uncertainty in the Power Spectra}
\label{sec:errors}
We estimate the uncertainty of the real parts of the power spectra from an estimate of their variance. To do this, we divide the data into four bins along the JD-axis. Using these four bins, we form six cross-power spectra from which we can compute three independent differences. Any true signal is cancelled in these differences, leaving us with three independent realisations of the noise (see \cite{closurelimits} and \cite{errorbar} for similar approaches). We then perform the averaging described in Section~\ref{sec:avg2} and obtain error bars by computing the root-mean-square (RMS) of the real parts of the three noise realisations. The errors thus obtained are themselves subject to uncertainties because of the small sample size. To decrease this uncertainty, we smooth the error bars in quadrature along the delay axis with a flat kernel of width 3. As a result, neighbouring power spectrum errors will be correlated, but instead we have effectively increased the sample size from 3 to 9. Averaging negative and positive $|\kappa_{||}|$-bins further increases the sample size to 18. The relative uncertainty of the resulting error bars is about 17\%.

The advantage of this method is that each $|\kappa_{||}|$-bin has its own error, which captures the variance due to thermal noise as well as systematic noise and accounts for cross-terms between the noise and any underlying signal. The subsequent smoothing, however, makes the assumption that neighbouring error bars have similar values. This assumption is accurate for thermal-like white noise, but can lead to biased estimates in regions where the noise-signal cross-terms start to dominate. Fortunately, we are interested in regions where these cross-terms are minimal and the errors are at most overestimated. 

\subsection{Estimating Upper Limits}
\label{sec:ul}
An upper limit $x^{ }_\mathrm{UL}$ on the closure phase power spectrum is implicitly given by the probability $\mathrm{Pr}\left(0 < \mu < x^{ }_\mathrm{UL}\right) = 1 - \alpha$, where $\mu$ is the true power. We choose $\alpha=0.05$ with which the upper limit defines a 95\% confidence interval. Assuming that the power spectrum data $x$ is drawn from a normal distribution $N(\mu, \sigma)$ with expectation $\mu$ and variance $\sigma$, we can apply Bayes theorem:
\begin{equation}
    \mathrm{Pr}\left(0 < \mu < x^{ }_\mathrm{UL}\right) = \frac{\int_0^{x^{ }_\mathrm{UL}}N(x|\mu, \sigma)\mathrm{d}\mu}{\int_0^{\infty} N(x|\mu, \sigma)\mathrm{d}\mu},
\end{equation}
where setting the lower integral limit to zero incorporates our prior knowledge of the true power spectrum and the denominator normalises the posterior probability. Computing the integrals, we find

\begin{equation}
    \mathrm{Pr}\left(0 < \mu < x^{ }_\mathrm{UL}\right) = \frac{\mathrm{erf}\left(\left(x^{ }_\mathrm{UL} - x\right)/\sqrt{2}\sigma\right) + \mathrm{erf}\left(x/\sqrt{2}\sigma\right)}{1 + \mathrm{erf}\left(x/\sqrt{2}\sigma\right)},
\end{equation}
where $\mathrm{erf}$ is the error function. Defining $E:=\mathrm{erf}\left(x/\sqrt{2}\sigma\right)$, we can solve for the upper limit

\begin{equation}
    x^{ }_\mathrm{UL} = \sqrt{2}\sigma  \left(\mathrm{erf}^{-1}(1 - (1+E)\alpha) + \mathrm{erf}^{-1}(E)\right).
\end{equation}
Although the closure phase delay power spectra do not initially follow a normal distribution, they will converge to normality as more data is averaged \citep[cf.][]{errorbar}. This is a consequence of the Central Limit Theorem. The degree to which the noise in our power spectra is normally distributed is investigated in Section~\ref{sec:results}. A similar derivation of this result can be found in the appendix of \cite{MWAII}.

\begin{figure*}
    \centering
    \includegraphics[width=0.9\linewidth]{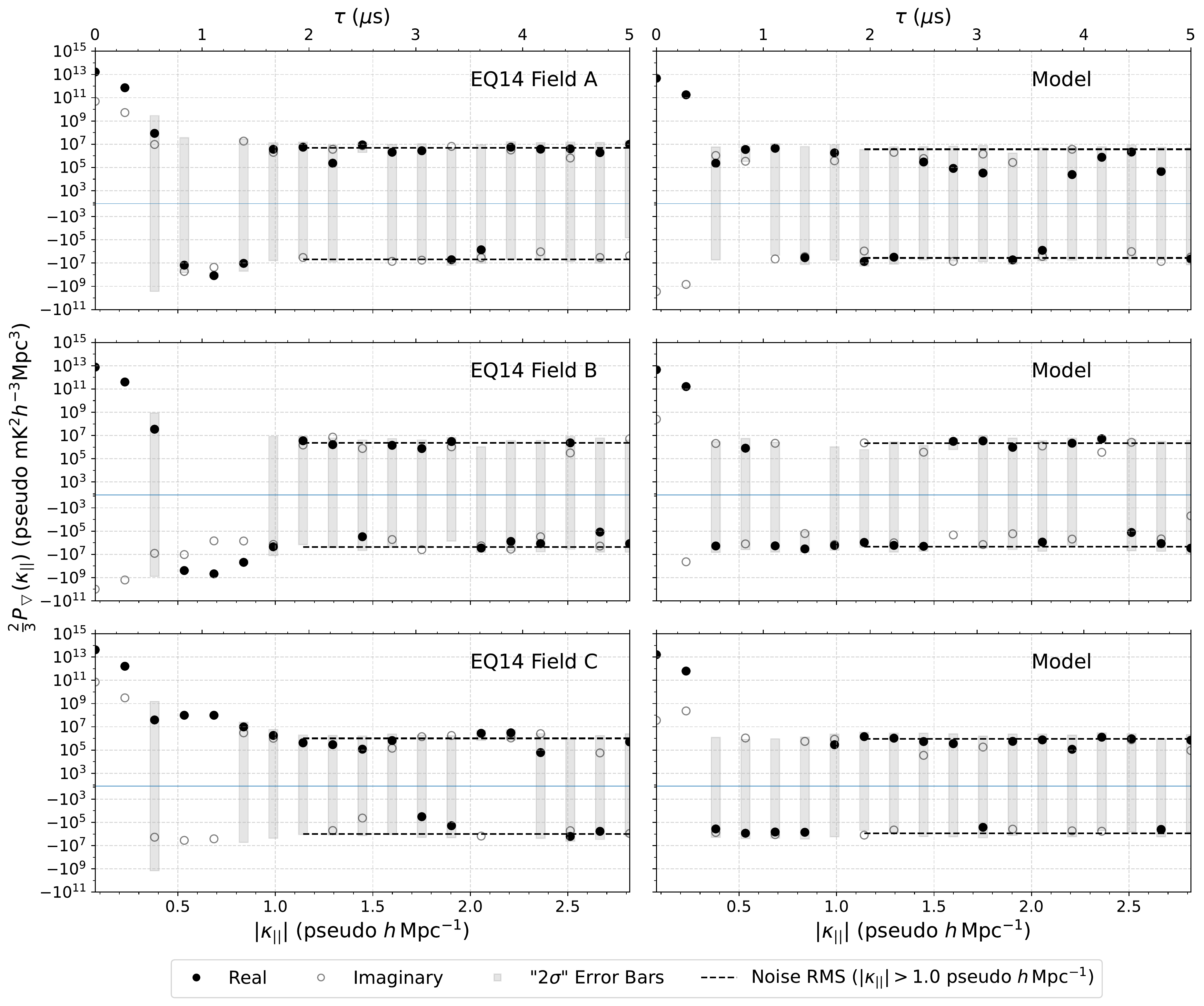}

    \caption{Closure phase delay power spectra for equilateral 14.6\,m triads and HERA-fields A, B,and C from data (left) and the corresponding models (right). The power is scaled by two-thirds to account for the fact that the the closure phase is formed by three different phases, which each, statistically, recover half the visibility fluctuations. The observed band is centred at $\sim$164\,MHz, which corresponds to a redshift of $\sim$7.7. The axes are given in "pseudo" units in order to discriminate between true cosmological scales and the approximate scales used in this analysis. In addition to the real parts of the cross-power spectra (black circles), we also show the imaginary parts (gray circles), the "$2\sigma$" error bars (gray boxes) and the root-mean-squares of the power spectrum errors at $|\kappa_{||}|>1.0$ "pseudo" $h\,\mathrm{Mpc}^{-1}$ (black dashed lines). The power axes are linear in the regions between $-10^3$ and $10^3$ "pseudo" mK$^2h^{-3}$Mpc$^3$ and logarithmic otherwise.}
    \label{fig:xps_eq14}
\end{figure*}
\begin{figure*}
    \centering
    \includegraphics[width=0.9\linewidth]{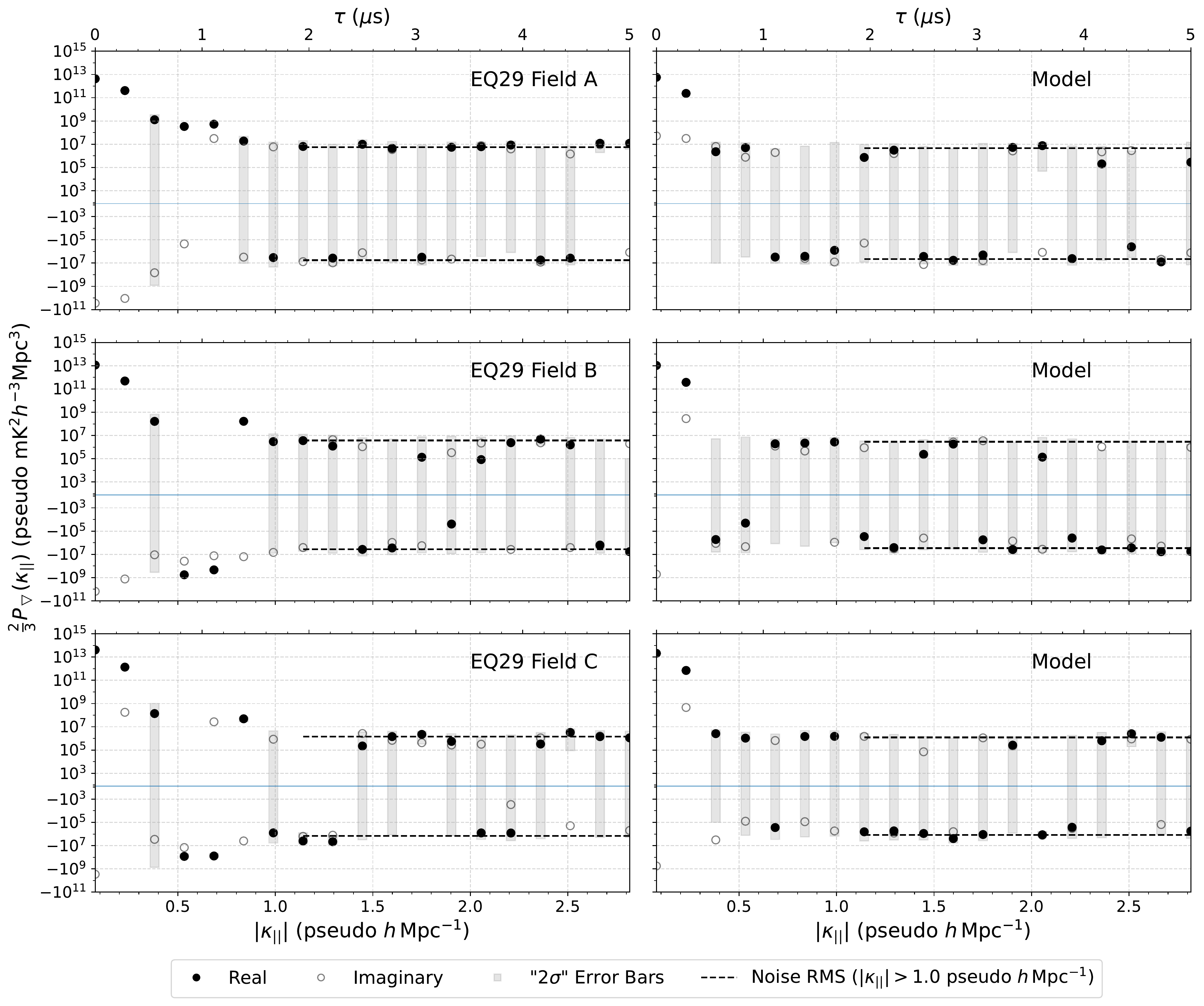}

    \caption{Same as figure~\ref{fig:xps_eq14} but for equilateral 29-metre triads.}
    \label{fig:xps_eq28}
\end{figure*}
\begin{figure*}
    \centering
    \includegraphics[width=0.9\linewidth]{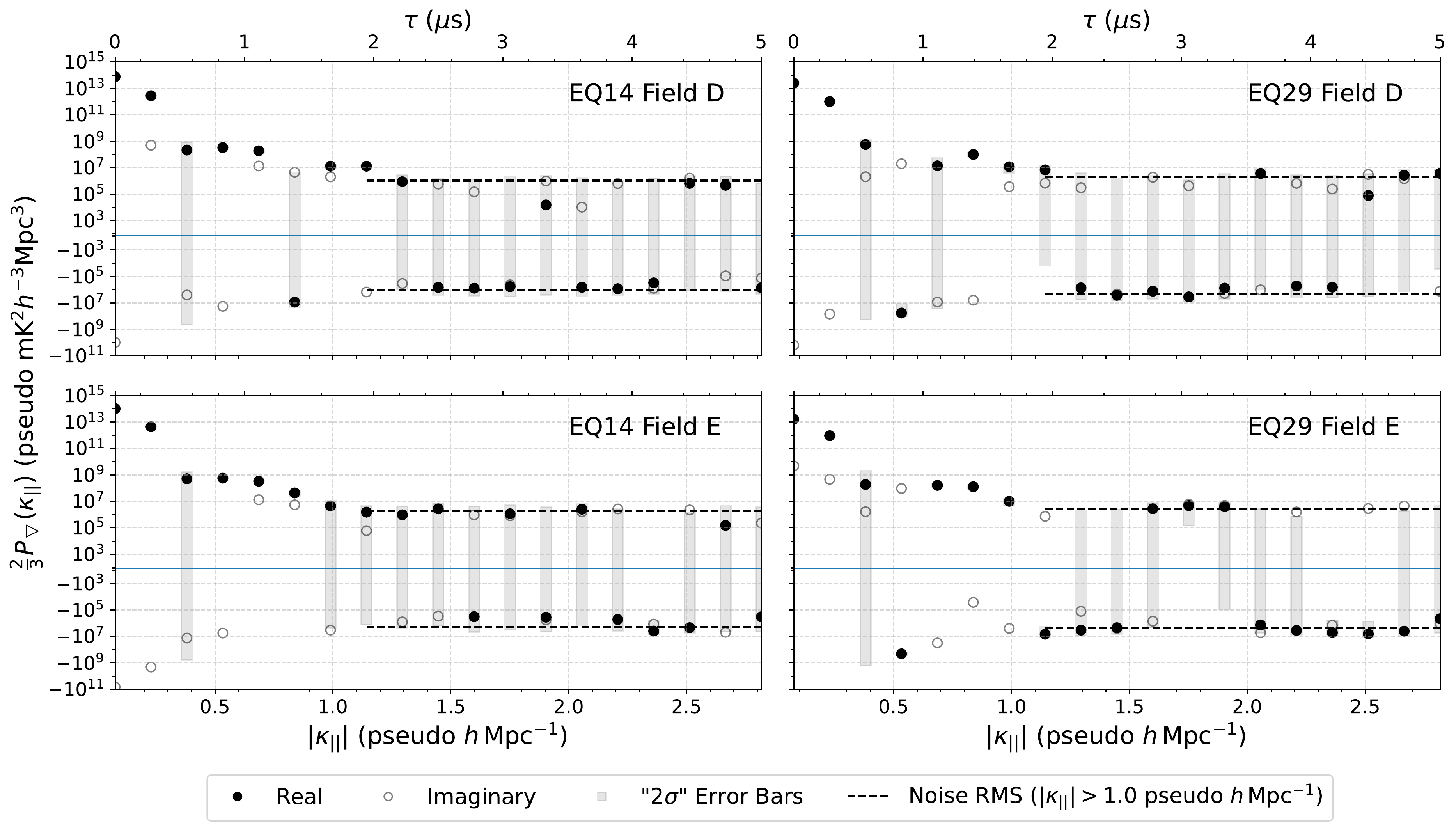}

    \caption{Closure phase delay power spectra of HERA-fields D and E. Unlike the other fields, these two fields cannot be modelled adequately. The left and right columns show the power spectra for equilateral 14.6\,m and 29.2\,m triads respectively. }
    \label{fig:xps_DE}
\end{figure*}

\section{Results}
\label{sec:results}

The final averaged power spectra for fields A, B and C are shown in Figure~\ref{fig:xps_eq14} (EQ14) and Figure~\ref{fig:xps_eq28} (EQ29), where the left and right columns show the power spectra obtained from data and the corresponding models respectively. The power spectra obtained on fields D and E are shown separately in Figure~\ref{fig:xps_DE}. Since fields D and E where not modelled, we show the power spectra obtained from EQ14 triads in the left column and those obtained from EQ29 triads in the right column. Note that the power spectra are scaled by a factor of 2/3 to account for the fact that the fluctuations in the closure phase enter through three visibilities and that, statistically, the phase only recovers half of the visibility fluctuations \citep[cf.][]{closuremaths}. The plots show the real part of the power spectra (black filled circles) as well as the imaginary part (gray circles). The latter should be a good proxy of the noise including signal-noise cross-terms and can be compared to the real parts to identify excess power above the noise level. Excess power in the imaginary part, on the other hand, would either indicate significant systematics-noise cross-terms or certain classes of systematics that vary in phase across nights or triads. The plots also show the $2\sigma$ error bars (gray boxes) and the RMS of the noise at $|\kappa_{||}| > 1.0$ "pseudo" $h$\,Mpc$^{-1}$ (black dashed lines). The power axes is in a symmetric log-scale with a linear region between $\pm10^3$ "pseudo" mK$^2h^{-3}$Mpc$^3$, allowing us to show positive and negative powers on the same plot.

We can distinguish between three regions in the power spectra. The first region is at $|\kappa_{||}|\lesssim 0.3$ "pseudo" $h\,$Mpc$^{-1}$ where the largest fraction of the power is concentrated. This power is attributed to the spectrally smooth foreground emissions and peaks between $10^{12}$ and $10^{14}$ "pseudo" mK$^2h^{-3}$Mpc$^3$ depending on the observed field. That is, fields containing strong emissions (e.g. Field E) peak higher than fields with weak emissions (e.g. Field B). Moreover, the power spectra of EQ29 peak lower than those of EQ14, because of the weaker response to large scale emissions on the sky. Since this analysis takes a foreground avoidance approach, we do not set any upper limits in this region of the power spectra.

The second region is characterised by peaks of amplitude $10^8$-$10^9$ "pseudo" mK$^2h^{-3}$Mpc$^3$, which are centred at delays of $1\,\mu\mathrm{s}$ and stretch out to delays of about $1.4\,\mu\mathrm{s}$. These peaks are not seen in the models and have uncertainties that are not consistent with thermal-like noise. In closure phase spectra, this effect appears as a spectral ripple with a period of $\sim$1\,MHz. It is not a multiplicative antenna based effect, as it would otherwise be eliminated in the closure phase. In fact, in a visibility based analysis \cite{systematics1}, \cite{xtalkmemo} and \citetalias{upperlimits2} find evidence that the ripple is a baseline-dependent systematic effect caused by over-the-air coupling between different array elements. Furthermore, the effect is found to vary slowly in time but strongly across baselines \citep[cf.][]{systematics1}. Both triad classes are equally affected by this systematic effect. In the averaged power spectra, the peaks take on negative as well as positive values, indicating a partial de-correlation between triads. This agrees with the findings of \cite{closurelimits} that the peaks are considerably suppressed when excluding the cross-power between identical triads (cf. Section~\ref{sec:avg2}). While the spectral ripple is fitted for and subtracted in the visibility processing \citep{systematics1}, we do not use filtered data to form closure phases. The subtraction violates closure properties and would have to be performed on raw data rather than averaged data, which would be computationally expensive. Filtering the systematic directly from the closure phase also has its problems. The filtering in \citep{systematics1} makes use of the fact that the systematic varies slowly in time, therefore having a fringe rate close to zero. However, the same applies to closure quantities, which can be shown to be invariant under the translation of the sky. Filtering the systematic from the closure phase would therefore also remove a large part of the cosmological signal. Hence, we take the more conservative approach and completely avoid the affected delay modes.

\begin{figure*}
    \centering
    \includegraphics[width=0.9\linewidth]{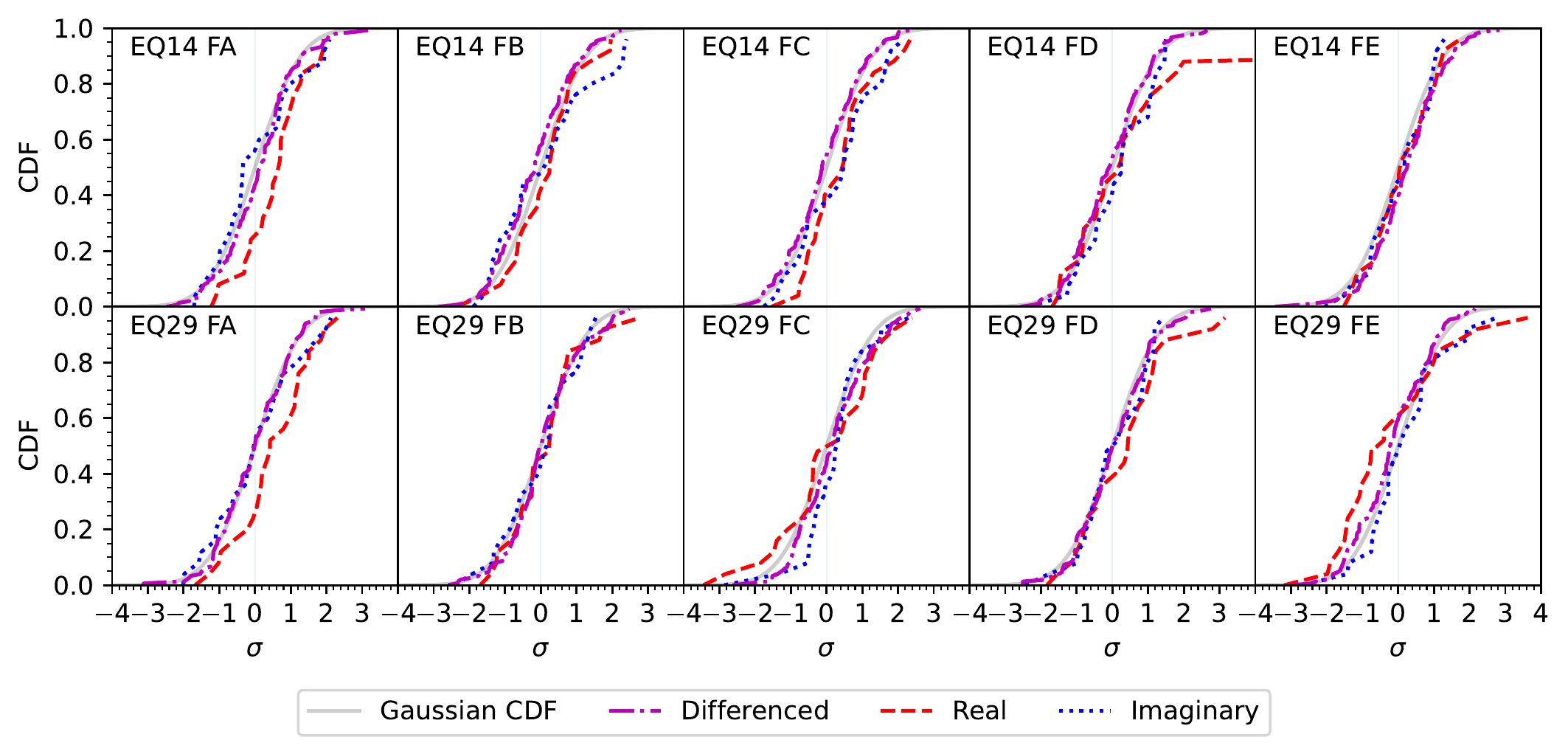}
    
    \caption{The cumulative distribution functions of the cross-power spectra at $\kappa_{||}>1.0$ "pseudo" $h\,\mathrm{Mpc}^{-1}$. The red dashed and the blue dotted lines show the CDFs of the real and imaginary parts of the power spectra respectively, while the magenta dash dotted line shows the CDF of the noise realisation obtained by differencing closure phases with the same underlying sky signal. As a reference, we also plot a Guassian CDF with the same variance as the noise (gray solid line). The abscissa is given in units of standard deviations of the differenced closure phases.}
    \label{fig:cdf}
\end{figure*}

The third region at $\tau \gtrsim 1.4$~$\mu\mathrm{s}$ or equivalently $|\kappa_{||}| \gtrsim 0.8$ "pseudo" $h$\,Mpc$^{-1}$ is dominated by noise, meaning that the power is comparable to its overall variance. In other words, the error bars mostly cross $P_\triangledown = 0$. Note that by "noise" we mean thermal-like noise as well as non-thermal-like noise such as RFI and instrumental effects. For thermal-like noise the high-delay region of the power spectrum should fluctuate randomly around zero. However, some of the power spectra have a tendency towards positive values in their real parts. To see this more clearly, we investigate the cumulative distribution functions (CDF) of the the real and imaginary parts of the power spectrum at $|\kappa_{||}|>1$ "pseudo" $h$\,Mpc$^{-1}$ as well as the combined real and imaginary parts of the differenced power spectra $P_\mathrm{Diff}$ used to obtain error bars (see Section~\ref{sec:errors}). We use the latter as a proxy for the noise. The CDFs are plotted in Figure~\ref{fig:cdf} together with the CDF of a Gaussian distribution (gray solid line) with the same variance as $P_\mathrm{Diff}$. Note that the CDFs of the differenced power spectra (magenta dashed line) are in good agreement with the Gaussian CDFs, which is a consequence of averaging many power spectra together (Central Limit Theorem). To quantify the consistency between the different CDFs, we use two statistical tests, the Shapiro-Wilk \citep[SW,][]{shapiro-wilk} and the Anderson-Darling \citep[AD,][]{anderson-darling} test, the results of which are shown in Table~\ref{tab:tests}. We use the former to test the null-hypothesis $H_\mathrm{N}$ that the differenced power spectra are drawn from a normal distribution. In all cases the SW test fails to reject $H_\mathrm{N}$ at the $5\%$ level. We use the AD test to test the null-hypothesis $H_\mathrm{N}$ that the noise and the real or imaginary parts of the power spectra at $|\kappa_{||}|>1$ "pseudo" $h$\,Mpc$^{-1}$ follow the same distribution. For fields A, B and C, we also use the AD test to test the data against the corresponding model at $|\kappa_{||}|>1$ "pseudo" $h$\,Mpc$^{-1}$. Unlike the SW test, the AD test does not make any assumptions about the shape of the underlying distributions. For the imaginary parts, the AD test fails in all cases to reject $H_\mathrm{N}$ at the $5\%$ level. For the real parts, $H_\mathrm{N}$ is rejected for the power spectrum of EQ29 on Field A, and not rejected otherwise. The test of the data against the models, on the other hand, fails to reject $H_\mathrm{N}$ for all power spectra except that of EQ14 on Field A. The CDFs of the real parts of Field A show a shift towards positive values, which is consistent with a 'detection' of a signal. This signal is of unknown origin and could be due to a variety of different effects such as RFI or digital artefacts. Since Field A is the least sensitive field, it should be less affected by low-level RFI. On the other hand, it is covered by fewer nights, thus making the median-averaging across nights less effective at rejecting RFI. 

Although not rejected by the AD test at a $5\%$ level, the CDFs of the real parts of Field D extend towards high positive values. Looking at the power spectrum, we see a peak of excess power at delays of about $2\,\mu\mathrm{s}$. \citetalias{upperlimits} identify a similar feature in the visibility delay spectrum and trace it to the polarised emissions of the pulsar PSR J0742-2822 \citep{pulsar}. The polarisation direction is rotated as the radiation passes through magnetic fields. This effect, known as Faraday rotation, is frequency dependent, thus leaving an imprint on certain delay modes in the power spectrum. In the visibility analysis, the Faraday effect can be suppressed below the current noise level by forming pseudo Stokes~I visibilities \citepalias{upperlimits}. This is not possible for the closure phase approach, as each polarisation will have independent gains, meaning that the closure phase of a pseudo Stokes~I visibility would loose its desirable properties. Consequently, we cannot reliably interpret fields with strong highly-polarised sources at high rotation measure. 

The real parts of the power spectra on Field C and of EQ29 on Field E also have a somewhat higher AD statistic $A^2$ compared with other power spectra, albeit not high enough to be rejected at the $5\%$ level. In the CDFs of EQ29 Fields C and E, we see that both have an extended tail towards negative values, which could indicate the presence of a systematic effect that is uncorrelated across triads or nights. Moreover, since we do not see the same effect in the power spectra of the EQ14 triads, this demonstrates that the two triad classes are affected differently by systematic effects.

Comparing the noise RMS at $|\kappa_{||}|>1$ "pseudo" $h$\,Mpc$^{-1}$ between the data and the models, we find that the RMS's of the models are on average a factor of $\sim$1.2 lower than those of the data. Possible explanations of this discrepancy are the presence of non-thermal effects (e.g. RFI), an underestimate of the system temperature, or other model inaccuracies. As expected, we also find that the noise RMS of EQ14 power spectra are lower than those of EQ29 power spectra, since there are fewer triads in EQ29 that can be averaged.

\begin{figure}
    \centering
    \includegraphics[width=\linewidth]{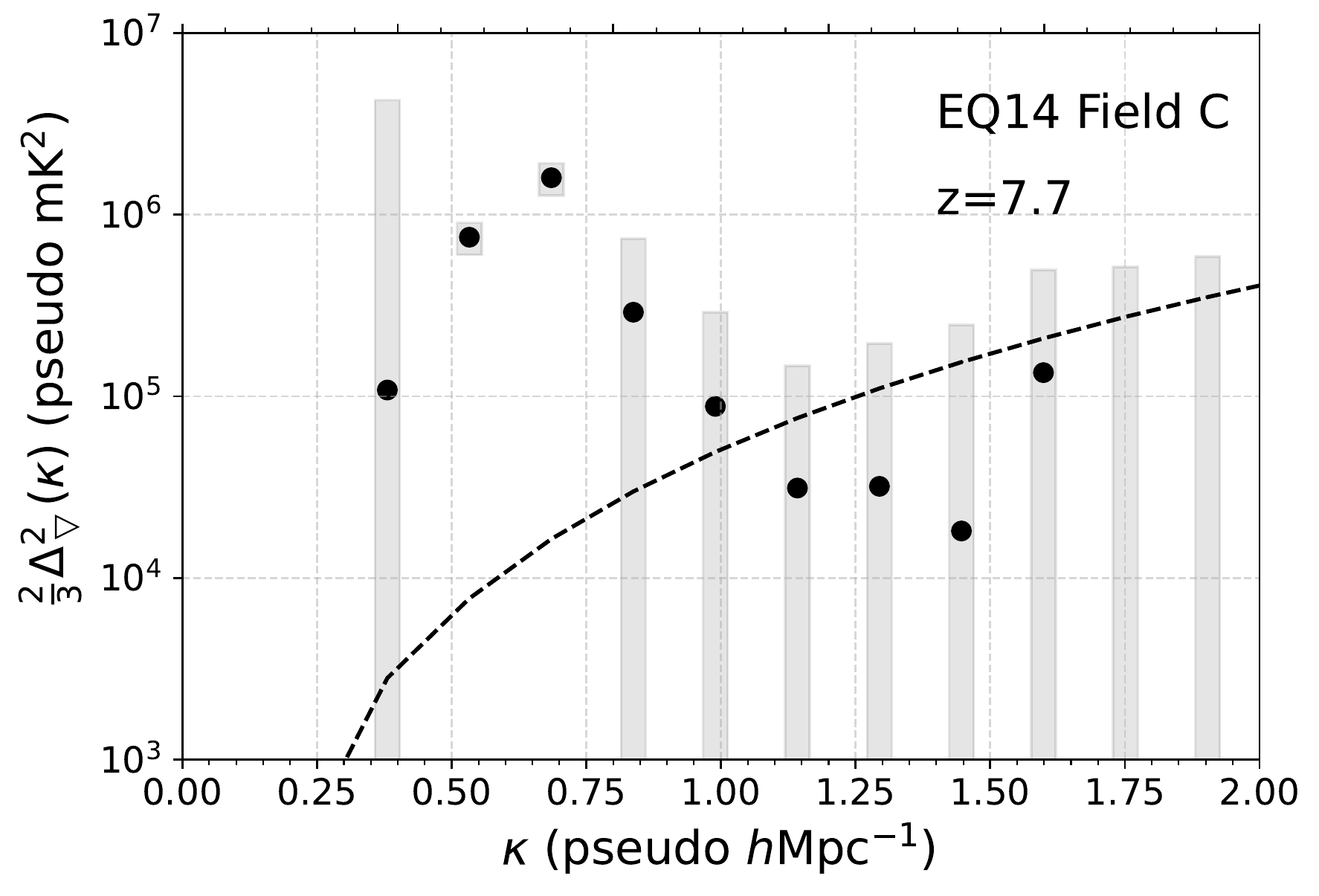}
    
    \caption{The closure phase power spectrum of EQ14 Field C in units of mK$^2$ (cf. Figure~\ref{fig:xps_eq14}). Data points with negative powers are only shown by their error bars (gray boxes) on this plot. The dashed line indicates the RMS of the power spectrum errors at $|\kappa_{||}|>1.0$ "pseudo" $h\,\mathrm{Mpc}^{-1}$.}
    \label{fig:ul}
\end{figure}

Despite the presence of non-thermal like effects in some of the power spectra, we can use the Bayesian framework described in Section~\ref{sec:ul} to set upper limits on the cosmological 21\,cm signal. Here, we provide the limits for our deepest field and triad class, Field C and EQ14. Figure~\ref{fig:ul} shows the power spectrum in units of "pseudo" mK$^2$ and Table~\ref{tab:ul} shows the associated upper limits at different $\kappa$-modes. The strongest noise-limited upper limit is (372)$^2$ "pseudo" mK$^2$ at 1.14 "pseudo" $h$\,Mpc$^{-1}$. We re-emphasise that this limit should only be interpreted as approximations to the physical distance and brightness scales of the conventional 21\,cm power spectrum \citep[cf.][]{closuremaths}.

\begin{table*}
    \centering
    \caption{Summary of the results from statistical tests performed on the cross-power spectra at $\kappa_{||}>1.0$ "pseudo" $h\,\mathrm{Mpc}^{-1}$. The variables $W$ and $A^2$ denote the test statistics of the Shapiro-Wilk and Anderson-Darling test respectively. We set our critical value for rejecting the null-hypothesis $H_\mathrm{N}$ at 5\%. Rows for which $H_\mathrm{N}$ is rejected are shaded magenta while rows for which the test fails to reject $H_\mathrm{N}$ are shaded yellow. See the table notes for further elaboration.}
    \renewcommand{\arraystretch}{1.2} 
    
    \setlength{\tabcolsep}{0.39cm}
    \definecolor{myyellow}{HSB}{40, 20, 240}
    \definecolor{mymagenta}{HSB}{200, 20, 240}
    \begin{threeparttable}
    \begin{tabular*}{\linewidth}{l|ccc|cc|cc|cc}
        \hline\hline
        \multicolumn{1}{c|}{Test} & \multicolumn{3}{c|}{Shapiro-Wilk$^1$} & \multicolumn{2}{c|}{Anderson-Darling$^2$} & \multicolumn{2}{c|}{Anderson-Darling$^2$} & \multicolumn{2}{c}{Anderson-Darling$^2$}\\ 
        \multicolumn{1}{c|}{Samples} & \multicolumn{3}{c|}{$P_\mathrm{Diff}$} & \multicolumn{2}{c|}{$P_\mathrm{Diff}$ and $\Re \left\{P_\triangledown\right\}$} & \multicolumn{2}{c|}{$P_\mathrm{Diff}$ and $\Im \left\{P_\triangledown \right\}$} &
        \multicolumn{2}{c}{$\Re \left\{P_\triangledown \right\}$ model and data}\\ \hline
        \multicolumn{1}{c|}{Name} & $W$ & $p$-value & reject $H_\mathrm{N}$ & $A^2$ & reject $H_\mathrm{N}$ & $A^2$ & reject $H_\mathrm{N}$ & $A^2$ & reject $H_\mathrm{N}$\\ 
EQ14 Field A & \cellcolor{myyellow} 0.992 & \cellcolor{myyellow} 0.605 & \cellcolor{myyellow} False & \cellcolor{myyellow} 1.889 & \cellcolor{myyellow} False & \cellcolor{myyellow} -0.043 & \cellcolor{myyellow} False & \cellcolor{mymagenta} 4.877 & \cellcolor{mymagenta} True\\
EQ14 Field B & \cellcolor{myyellow} 0.994 & \cellcolor{myyellow} 0.762 & \cellcolor{myyellow} False & \cellcolor{myyellow} -0.062 & \cellcolor{myyellow} False & \cellcolor{myyellow} 0.425 & \cellcolor{myyellow} False & \cellcolor{myyellow} -0.976 & \cellcolor{myyellow} False\\
EQ14 Field C & \cellcolor{myyellow} 0.994 & \cellcolor{myyellow} 0.842 & \cellcolor{myyellow} False & \cellcolor{myyellow} 0.971 & \cellcolor{myyellow} False & \cellcolor{myyellow} 0.560 & \cellcolor{myyellow} False & \cellcolor{myyellow} 0.140 & \cellcolor{myyellow} False\\
EQ14 Field D & \cellcolor{myyellow} 0.993 & \cellcolor{myyellow} 0.627 & \cellcolor{myyellow} False & \cellcolor{myyellow} -0.051 & \cellcolor{myyellow} False & \cellcolor{myyellow} -0.474 & \cellcolor{myyellow} False &  - &  -\\
EQ14 Field E & \cellcolor{myyellow} 0.989 & \cellcolor{myyellow} 0.283 & \cellcolor{myyellow} False & \cellcolor{myyellow} -0.627 & \cellcolor{myyellow} False & \cellcolor{myyellow} -0.413 & \cellcolor{myyellow} False &  - &  -\\
EQ28 Field A & \cellcolor{myyellow} 0.990 & \cellcolor{myyellow} 0.342 & \cellcolor{myyellow} False & \cellcolor{mymagenta} 2.319 & \cellcolor{mymagenta} True & \cellcolor{myyellow} -0.523 & \cellcolor{myyellow} False & \cellcolor{myyellow} 1.787 & \cellcolor{myyellow} False\\
EQ28 Field B & \cellcolor{myyellow} 0.985 & \cellcolor{myyellow} 0.098 & \cellcolor{myyellow} False & \cellcolor{myyellow} -0.904 & \cellcolor{myyellow} False & \cellcolor{myyellow} -0.605 & \cellcolor{myyellow} False & \cellcolor{myyellow} 0.883 & \cellcolor{myyellow} False\\
EQ28 Field C & \cellcolor{myyellow} 0.991 & \cellcolor{myyellow} 0.474 & \cellcolor{myyellow} False & \cellcolor{myyellow} 0.450 & \cellcolor{myyellow} False & \cellcolor{myyellow} -0.399 & \cellcolor{myyellow} False & \cellcolor{myyellow} -0.106 & \cellcolor{myyellow} False\\
EQ28 Field D & \cellcolor{myyellow} 0.993 & \cellcolor{myyellow} 0.684 & \cellcolor{myyellow} False & \cellcolor{myyellow} -0.172 & \cellcolor{myyellow} False & \cellcolor{myyellow} -0.912 & \cellcolor{myyellow} False &  - &  -\\
EQ28 Field E & \cellcolor{myyellow} 0.993 & \cellcolor{myyellow} 0.661 & \cellcolor{myyellow} False & \cellcolor{myyellow} 0.936 & \cellcolor{myyellow} False & \cellcolor{myyellow} -0.185 & \cellcolor{myyellow} False &  - &  -\\

 \hline\hline
    \end{tabular*}
    \begin{tablenotes}
    \item[1] The null hypothesis $H_\mathrm{N}$ of the Shapiro-Wilk test is that $P_\mathrm{Diff}$ is drawn from a normal distribution. 
    \item[2] The null hypothesis $H_\mathrm{N}$ of the two sample Anderson-Darling test is that two samples are drawn from the same distributions. Here, we test $P_\mathrm{Diff}$ against $\Re\left\{P_\triangledown\right\}$ and $\Im\left\{P_\triangledown\right\}$, and the models against the data. The critical values of the test statistic $A^2$ are 0.325, 1.226, 1.961, 2.718, 3.752, 4.592 and 6.546 at 25\%, 10\%, 5\%, 2.5\%, 1\%, 0.5\% and 0.1\% significant levels respectively.
    \end{tablenotes}
    \end{threeparttable}
    \label{tab:tests}
\end{table*}

\begin{table}
\renewcommand{\arraystretch}{1.2} 
\caption{The upper limits $(2/3)\Delta^\mathrm{2}_{\triangledown\mathrm{\,UL}}$ at 95\% confidence and the standard deviation $\sigma$ of the closure phase delay power spectrum obtained from EQ14 triads on Field C between 0.38 and 1.45 "pseudo" $h$\,Mpc$^{-1}$}
\setlength{\tabcolsep}{0.5cm}
    \centering
    \begin{tabular}{c|cc}
    \hline\hline
    $\kappa$ & $(2/3)\Delta^\mathrm{2}_{\triangledown\mathrm{\,UL}}$ & $\sigma$ \\
"pseudo" $h$\,Mpc$^{-1}$ & "pseudo" mK$^2$ & "pseudo" mK$^2$\\\hline
0.38 & (2058)$^2$ & (1443)$^2$\\
0.53 & (937)$^2$ & (274)$^2$\\
0.69 & (1367)$^2$ & (401)$^2$\\
0.84 & (824)$^2$ & (473)$^2$\\
0.99 & (518)$^2$ & (317)$^2$\\
1.14 & (372)$^2$ & (241)$^2$\\
1.29 & (431)$^2$ & (285)$^2$\\
1.45 & (491)$^2$ & (338)$^2$\\
\hline\hline
    \end{tabular}
    \label{tab:ul}
\end{table}

\section{Summary}
\label{sec:summary}
We present closure phase delay power spectra using data from a full season of HERA Phase~I observing. The data was observed over 94 unflagged nights using 48 antennas. We show power spectra for two triad classes, equilateral 14.6 and 29.2-metre triads, and five separate LST ranges. Using our most sensitive field, Field C, which covers 2.25\,h of LST centered at 5.125\,h, we provide upper limits on the closure phase power spectrum of equilateral 14.2-metre triads and find a noise-limited upper limit of (372)$^2$ "pseudo" mK$^2$ at 1.14 "pseudo" $h$\,Mpc$^{-1}$. Comparing the power spectrum RMS at $\kappa>1.0$ "pseudo" $h$\,Mpc$^{-1}$ with that of \cite{closurelimits} at $\kappa>0.85$ "pseudo" $h$\,Mpc$^{-1}$, we find an improvement in sensitivity by a factor of $\sim$26. The limits reported here are only approximately related to the true distance and brightness scales of the redshifted 21\,cm power spectrum, which is why they are given in "pseudo" units. A more careful interpretation of results obtained with the closure phase requires detailed forward-modelling of the sky signals. However, making use of the approximation with "pseudo" units, we find that our most sensitive noise-limited upper limit is a factor of $\sim$5 above that of \citetalias{upperlimits2} at $k$=1.16\,$h$\,Mpc$^{-1}$.

\begin{figure}
    \centering
    \includegraphics[width=\linewidth]{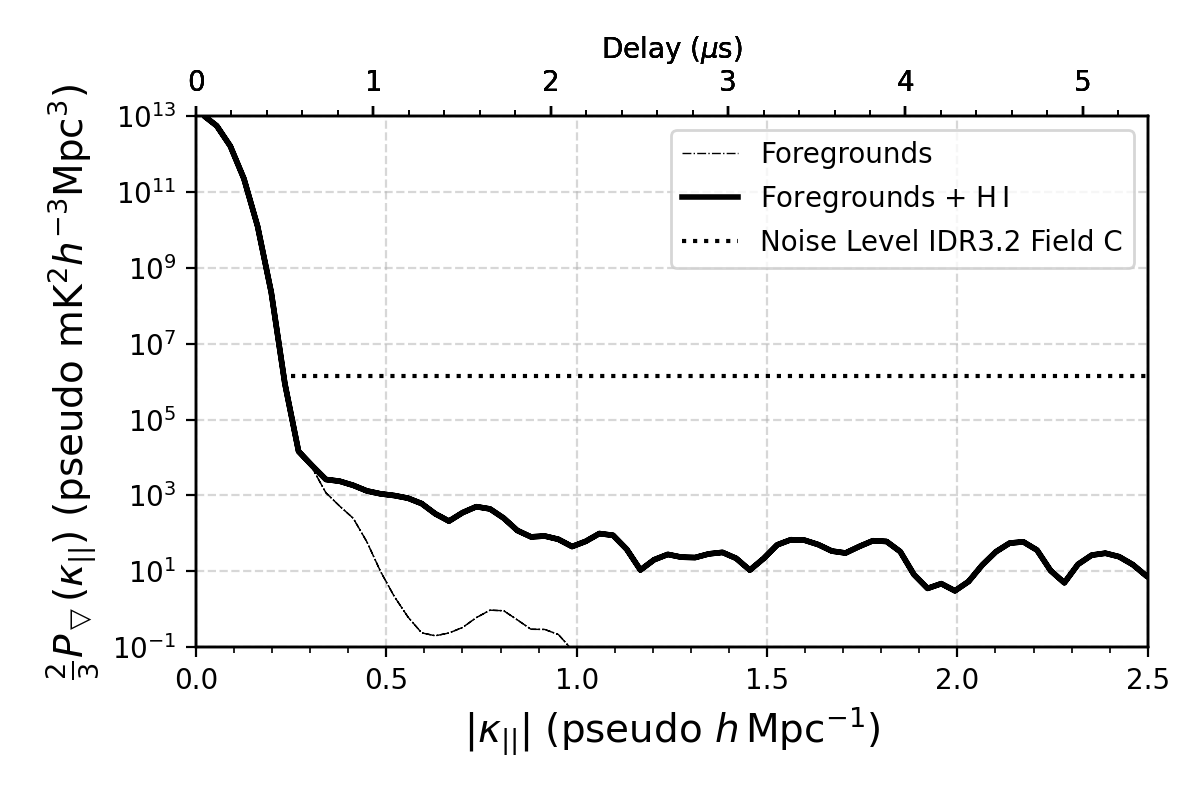}
    \caption{A simulated delay power spectrum of the frequency band 152.25-167.97\,MHz ($z\sim7.9$). The dashed line is the power spectrum of a sky with foregrounds only (GLEAM sources at 5\,h LST) and the solid line is that of a sky including a cosmological \textsc{H\,i} signal (Faint Galaxies, 21cmFast). While the foreground dominate the low $|\kappa_{||}|$-modes, the \textsc{H\,i} signal dominates at $|\kappa_{||}|>0.25$. Also shown is the noise level at $|\kappa_{||}|>1.0$ of our most sensitive field (hoizontal dotted line). An improvement in sensitivity of at least two orders of magnitude is required for a hypothetical detection of this commonly used EoR model.}
    \label{fig:sim}
\end{figure}

As can be seen in Figure~\ref{fig:sim}, the sensitivity of the power spectrum needs to be improved by at least two orders of magnitude to achieve a hypothetical detection of the EoR model used here (see Section~\ref{sec:modelling}). When fully operational, HERA will have 320 antennas which can be used for the closure phase analysis. This amounts to $\sim$7 times more nominally redundant triads than are used in this analysis, which can be combined coherently to improve the sensitivity by the same factor. Moreover, HERA will observe all year round for $\sim$12 hours per night. Combining different fields incoherently could potentially double the sensitivity ($\sim$4 more data in LST). Repeated nights can also be combined coherently. Taken together, it should therefore be possible to improve the power spectrum sensitivity by a factor greater than $10^2$ within the observing horizon of HERA ($\sim$5 years). This should be sufficient for a detection of commonly assumed 21\,cm signals at $\kappa\sim0.4\,h$\,Mpc$^{-1}$ provided that we are not limited by systematic effects.

There are some limitations to the analysis as it is presented here. Unlike the visibility analysis, we are not able to model and subtract any baseline-dependent systematic effects such as the 1\,$\mu$s spectral ripple without loosing the desirable properties of the closure phase. Consequently, we do not have access to the lowest $\kappa$-modes which, otherwise, are expected to be most sensitive to the cosmological signal. The new HERA system uses fibre optic transmission lines, which may eliminate the 1\,$\mu$s ripple and allow to access lower $\kappa$-modes.

A further limitation is that we do not combine different triad classes in our analysis. While in the visibility analysis many different baselines can be averaged by the method of so-called spherical averaging, there is no analogous method to do this for different triad classes. This is especially true for non-equilateral triangle shapes such as isosceles, scalene or linear triads that are formed by baselines of different lengths. Including all these triad classes would lead to a considerable improvement in sensitivity, but as long as there is no physically motivated way to combine them to a single power spectrum, the classes will have to be analysed separately from one another. In other words, the power spectra of the different triad classes need to be considered as separate measurements, which, combined and in conjunction with forward-modelling, can be used to infer the presence of a signal. This combined analysis may therefore open up a pathway to exploiting the full sensitivity of the array.

Lastly, we find that Faraday rotated emissions may contaminate EoR window of the closure phase delay power spectrum. This effect could be mitigated by forming the closure phase from pseudo Stokes~I visibilities. However, the antenna based gains are generally independent between different polarisations, meaning that we would loose some of the advantages of the closure phase approach. To completely bypass the problem of Faraday rotation, one would either have to directly measure Stokes~I (i.e. by using circular antenna feeds if the emissions are unpolarised) or resort to polarisation independent closure quantities such as the closure trace \citep{closuretrace}. The implications of using the latter have yet to be investigated in detail. For HERA, a closure phase based approach will need to avoid regions containing strong pulsars such as the galactic plane (e.g. Field D).

Despite these limitations, the closure phase analysis still retains advantages over the standard approach. Most importantly, our analysis is independent of multiplicative antenna-based effects, which allows us to bypass conventional calibration. As a result, we require considerably fewer analysis steps and expect fewer errors and systematic effects to be introduced in the data processing. 
%Using robust averaging and avoidance methods, we sidestep some of the sophisticated and computationally expensive methods that have been devised to prevent RFI and instrumental systematic, such as the $1\,\mu\mathrm{s}$ ripple, from contaminating the power spectrum \citep[e.g.][]{systematics1, systematics2, upperlimits}. 
%The analysis pipeline is therefore much simpler, easy to verify and, contrary to the visibility analysis, does not contain any non-linear processing steps which can lead to signal losses.

The closure phase analysis provides an alternative and independent method by which the 21\,cm signal during the EoR can be searched for. Initially, the prime objective is a first detection of the signal. The interpretation of a detected signal will require extensive forward-modelling of the sky, which includes the 21\,cm signal as well as foregrounds, as the closure phase is a higher order (non-linear) interferometric quantity. Future work will explore the possibilities of inferring astrophysical properties of the IGM from the closure phase delay power spectrum. 

\section*{Acknowledgements}
The software developed for this analysis uses Python and the publicly-accessible and open-sourced Python packages Numpy \citep{numpy}, SciPy \citep{scipy}, Astropy \citep{astropy} and Matplotlib \citep{matplotlib}.

This material is based upon work supported by the National Science Foundation under grants \#1636646 and \#1836019 and institutional support from the HERA collaboration partners. This research is funded in part by the Gordon and Betty Moore Foundation through Grant GBMF5212 to the Massachusetts Institute of Technology. HERA is hosted by the South African Radio Astronomy Observatory, which is a facility of the National Research Foundation, an agency of the Department of Science and Innovation. 

P.M. Keller is funded by the Institute of Astronomy and Physics Department of the University of Cambridge via the Isaac Newton Studentship. N. Kern gratefully acknowledges support from the MIT Pappalardo fellowship. A. Liu acknowledges support from the New Frontiers in Research Fund Exploration grant program, the Canadian Institute for Advanced Research (CIFAR) Azrieli Global Scholars program, a Natural Sciences and Engineering Research Council of Canada (NSERC) Discovery Grant and a Discovery Launch Supplement, the Sloan Research Fellowship, and the William Dawson Scholarship at McGill.

%%%%%%%%%%%%%%%%%%%%%%%%%%%%%%%%%%%%%%%%%%%%%%%%%%
\section*{Data Availability}
The code used for the analysis is publicly available at \url{https://github.com/pm-keller/hera_closure} and \url{https://github.com/pm-keller/closure_sim}. The former repository also contains the data shown in Figures~\ref{fig:xps_eq14},~\ref{fig:xps_eq28},~\ref{fig:xps_DE} and~\ref{fig:ul} and Table~\ref{tab:ul} in a machine readable form. 



\bibliographystyle{mnras}
\bibliography{main.bib} % if your bibtex file is called example.bib

\begin{thebibliography}{}
\makeatletter
\relax
\def\mn@urlcharsother{\let\do\@makeother \do\$\do\&\do\#\do\^\do\_\do\%\do\~}
\def\mn@doi{\begingroup\mn@urlcharsother \@ifnextchar [ {\mn@doi@}
  {\mn@doi@[]}}
\def\mn@doi@[#1]#2{\def\@tempa{#1}\ifx\@tempa\@empty \href
  {http://dx.doi.org/#2} {doi:#2}\else \href {http://dx.doi.org/#2} {#1}\fi
  \endgroup}
\def\mn@eprint#1#2{\mn@eprint@#1:#2::\@nil}
\def\mn@eprint@arXiv#1{\href {http://arxiv.org/abs/#1} {{\tt arXiv:#1}}}
\def\mn@eprint@dblp#1{\href {http://dblp.uni-trier.de/rec/bibtex/#1.xml}
  {dblp:#1}}
\def\mn@eprint@#1:#2:#3:#4\@nil{\def\@tempa {#1}\def\@tempb {#2}\def\@tempc
  {#3}\ifx \@tempc \@empty \let \@tempc \@tempb \let \@tempb \@tempa \fi \ifx
  \@tempb \@empty \def\@tempb {arXiv}\fi \@ifundefined
  {mn@eprint@\@tempb}{\@tempb:\@tempc}{\expandafter \expandafter \csname
  mn@eprint@\@tempb\endcsname \expandafter{\@tempc}}}

\bibitem[\protect\citeauthoryear{{Aguirre} et~al.,}{{Aguirre}
  et~al.}{2022}]{HERAvalidation}
{Aguirre} J.~E.,  et~al., 2022, \mn@doi [\apj] {10.3847/1538-4357/ac32cd},
  \href {https://ui.adsabs.harvard.edu/abs/2022ApJ...924...85A} {924, 85}

\bibitem[\protect\citeauthoryear{Anderson \& Darling}{Anderson \&
  Darling}{1952}]{anderson-darling}
Anderson T.~W.,  Darling D.~A.,  1952, \mn@doi [The Annals of Mathematical
  Statistics] {10.1214/aoms/1177729437}, 23, 193

\bibitem[\protect\citeauthoryear{{Astropy Collaboration} et~al.,}{{Astropy
  Collaboration} et~al.}{2022}]{astropy}
{Astropy Collaboration} et~al., 2022, \mn@doi [apj] {10.3847/1538-4357/ac7c74},
  \href {https://ui.adsabs.harvard.edu/abs/2022ApJ...935..167A} {935, 167}

\bibitem[\protect\citeauthoryear{{Barry}, {Hazelton}, {Sullivan}, {Morales}  \&
  {Pober}}{{Barry} et~al.}{2016}]{barry2016calibration}
{Barry} N.,  {Hazelton} B.,  {Sullivan} I.,  {Morales} M.~F.,   {Pober} J.~C.,
  2016, \mn@doi [\mnras] {10.1093/mnras/stw1380}, \href
  {https://ui.adsabs.harvard.edu/abs/2016MNRAS.461.3135B} {461, 3135}

\bibitem[\protect\citeauthoryear{{Barry} et~al.,}{{Barry}
  et~al.}{2019}]{barry2019mwa}
{Barry} N.,  et~al., 2019, \mn@doi [\apj] {10.3847/1538-4357/ab40a8}, \href
  {https://ui.adsabs.harvard.edu/abs/2019ApJ...884....1B} {884, 1}

\bibitem[\protect\citeauthoryear{{Beardsley} et~al.,}{{Beardsley}
  et~al.}{2016}]{beardsley2016mwa}
{Beardsley} A.~P.,  et~al., 2016, \mn@doi [\apj] {10.3847/1538-4357/833/1/102},
  \href {https://ui.adsabs.harvard.edu/abs/2016ApJ...833..102B} {833, 102}

\bibitem[\protect\citeauthoryear{Blackman \& Tukey}{Blackman \&
  Tukey}{1958}]{blackman-harris}
Blackman R.~B.,  Tukey J.~W.,  1958, \mn@doi [The Bell System Technical
  Journal] {10.1002/j.1538-7305.1958.tb03874.x}, 37, 185

\bibitem[\protect\citeauthoryear{{Broderick} \& {Pesce}}{{Broderick} \&
  {Pesce}}{2020}]{closuretrace}
{Broderick} A.~E.,  {Pesce} D.~W.,  2020, \mn@doi [\apj]
  {10.3847/1538-4357/abbd9d}, \href
  {https://ui.adsabs.harvard.edu/abs/2020ApJ...904..126B} {904, 126}

\bibitem[\protect\citeauthoryear{{Byrne} et~al.,}{{Byrne}
  et~al.}{2019}]{byrne2019limitations}
{Byrne} R.,  et~al., 2019, \mn@doi [\apj] {10.3847/1538-4357/ab107d}, \href
  {https://ui.adsabs.harvard.edu/abs/2019ApJ...875...70B} {875, 70}

\bibitem[\protect\citeauthoryear{{Byrne}, {Morales}, {Hazelton}  \&
  {Wilensky}}{{Byrne} et~al.}{2021}]{byrne2021unified}
{Byrne} R.,  {Morales} M.~F.,  {Hazelton} B.~J.,   {Wilensky} M.,  2021,
  \mn@doi [\mnras] {10.1093/mnras/stab647}, \href
  {https://ui.adsabs.harvard.edu/abs/2021MNRAS.503.2457B} {503, 2457}

\bibitem[\protect\citeauthoryear{{Carilli}, {Nikolic}, {Thyagarayan}  \&
  {Gale-Sides}}{{Carilli} et~al.}{2018}]{closureHERA}
{Carilli} C.~L.,  {Nikolic} B.,  {Thyagarayan} N.,   {Gale-Sides} K.,  2018,
  \mn@doi [Radio Science] {10.1029/2018RS006537}, \href
  {https://ui.adsabs.harvard.edu/abs/2018RaSc...53..845C} {53, 845}

\bibitem[\protect\citeauthoryear{{Cheng} et~al.,}{{Cheng}
  et~al.}{2018}]{cheng2018paper}
{Cheng} C.,  et~al., 2018, \mn@doi [\apj] {10.3847/1538-4357/aae833}, \href
  {https://ui.adsabs.harvard.edu/abs/2018ApJ...868...26C} {868, 26}

\bibitem[\protect\citeauthoryear{{Choudhuri}, {Bull}  \& {Garsden}}{{Choudhuri}
  et~al.}{2021}]{redundancy2}
{Choudhuri} S.,  {Bull} P.,   {Garsden} H.,  2021, \mn@doi [\mnras]
  {10.1093/mnras/stab1795}, \href
  {https://ui.adsabs.harvard.edu/abs/2021MNRAS.506.2066C} {506, 2066}

\bibitem[\protect\citeauthoryear{{Datta}, {Bowman}  \& {Carilli}}{{Datta}
  et~al.}{2010}]{datta}
{Datta} A.,  {Bowman} J.~D.,   {Carilli} C.~L.,  2010, \mn@doi [\apj]
  {10.1088/0004-637X/724/1/526}, \href
  {https://ui.adsabs.harvard.edu/abs/2010ApJ...724..526D} {724, 526}

\bibitem[\protect\citeauthoryear{Daubechies}{Daubechies}{1988}]{daubechies1988orthonormal}
Daubechies I.,  1988, \mn@doi [Communications on Pure and Applied Mathematics]
  {https://doi.org/10.1002/cpa.3160410705}, 41, 909

\bibitem[\protect\citeauthoryear{{DeBoer} et~al.,}{{DeBoer}
  et~al.}{2017}]{hera2017}
{DeBoer} D.~R.,  et~al., 2017, \mn@doi [\pasp]
  {10.1088/1538-3873/129/974/045001}, \href
  {https://ui.adsabs.harvard.edu/abs/2017PASP..129d5001D} {129, 045001}

\bibitem[\protect\citeauthoryear{Dillon}{Dillon}{2021}]{memo97}
Dillon J.~S.,  2021, HERA Memorandum

\bibitem[\protect\citeauthoryear{{Dillon} \& {Parsons}}{{Dillon} \&
  {Parsons}}{2016}]{redundantConfig}
{Dillon} J.~S.,  {Parsons} A.~R.,  2016, \mn@doi [\apj]
  {10.3847/0004-637X/826/2/181}, \href
  {https://ui.adsabs.harvard.edu/abs/2016ApJ...826..181D} {826, 181}

\bibitem[\protect\citeauthoryear{{Dillon} et~al.,}{{Dillon}
  et~al.}{2015}]{dillon2015mwa}
{Dillon} J.~S.,  et~al., 2015, \mn@doi [\prd] {10.1103/PhysRevD.91.123011},
  \href {https://ui.adsabs.harvard.edu/abs/2015PhRvD..91l3011D} {91, 123011}

\bibitem[\protect\citeauthoryear{{Dillon} et~al.,}{{Dillon}
  et~al.}{2020}]{redcal}
{Dillon} J.~S.,  et~al., 2020, \mn@doi [\mnras] {10.1093/mnras/staa3001}, \href
  {https://ui.adsabs.harvard.edu/abs/2020MNRAS.499.5840D} {499, 5840}

\bibitem[\protect\citeauthoryear{Dillon, Parsons  \& Kern}{Dillon
  et~al.}{2021}]{xtalkmemo}
Dillon J.~S.,  Parsons A.,   Kern N.,  2021, HERA Memorandum

\bibitem[\protect\citeauthoryear{{Eastwood} et~al.,}{{Eastwood}
  et~al.}{2019}]{lwa}
{Eastwood} M.~W.,  et~al., 2019, \mn@doi [\aj] {10.3847/1538-3881/ab2629},
  \href {https://ui.adsabs.harvard.edu/abs/2019AJ....158...84E} {158, 84}

\bibitem[\protect\citeauthoryear{{Ewall-Wice}, {Dillon}, {Liu}  \&
  {Hewitt}}{{Ewall-Wice} et~al.}{2017}]{ewall2017calibration}
{Ewall-Wice} A.,  {Dillon} J.~S.,  {Liu} A.,   {Hewitt} J.,  2017, \mn@doi
  [\mnras] {10.1093/mnras/stx1221}, \href
  {https://ui.adsabs.harvard.edu/abs/2017MNRAS.470.1849E} {470, 1849}

\bibitem[\protect\citeauthoryear{{Fagnoni} et~al.,}{{Fagnoni}
  et~al.}{2021}]{beam}
{Fagnoni} N.,  et~al., 2021, \mn@doi [\mnras] {10.1093/mnras/staa3268}, \href
  {https://ui.adsabs.harvard.edu/abs/2021MNRAS.500.1232F} {500, 1232}

\bibitem[\protect\citeauthoryear{{Furlanetto}}{{Furlanetto}}{2016}]{furlanetto2016}
{Furlanetto} S.~R.,  2016, in {Mesinger} A.,  ed.,  Astrophysics and Space
  Science Library Vol. 423, Understanding the Epoch of Cosmic Reionization:
  Challenges and Progress. p.~247 (\mn@eprint {arXiv} {1511.01131}),
  \mn@doi{10.1007/978-3-319-21957-8_9}

\bibitem[\protect\citeauthoryear{{Greig} \& {Mesinger}}{{Greig} \&
  {Mesinger}}{2017}]{globalhistory}
{Greig} B.,  {Mesinger} A.,  2017, \mn@doi [\mnras] {10.1093/mnras/stw3026},
  \href {https://ui.adsabs.harvard.edu/abs/2017MNRAS.465.4838G} {465, 4838}

\bibitem[\protect\citeauthoryear{{{HERA Collaboration}}}{{{HERA
  Collaboration}}}{2022a}]{upperlimits2}
{{HERA Collaboration}} 2022a, arXiv e-prints, \href
  {https://ui.adsabs.harvard.edu/abs/2022arXiv221004912T} {p. arXiv:2210.04912}

\bibitem[\protect\citeauthoryear{{HERA Collaboration}}{{HERA
  Collaboration}}{2022b}]{HERAtheory}
{HERA Collaboration} 2022b, \mn@doi [\apj] {10.3847/1538-4357/ac2ffc}, \href
  {https://ui.adsabs.harvard.edu/abs/2022ApJ...924...51A} {924, 51}

\bibitem[\protect\citeauthoryear{{HERA Collaboration}}{{HERA
  Collaboration}}{2022c}]{upperlimits}
{HERA Collaboration} 2022c, \mn@doi [\apj] {10.3847/1538-4357/ac1c78}, \href
  {https://ui.adsabs.harvard.edu/abs/2022ApJ...925..221A} {925, 221}

\bibitem[\protect\citeauthoryear{{Hunter}}{{Hunter}}{2007}]{matplotlib}
{Hunter} J.~D.,  2007, Computing in Science Engineering, 9, 90

\bibitem[\protect\citeauthoryear{{Hurley-Walker} et~al.,}{{Hurley-Walker}
  et~al.}{2017}]{gleam}
{Hurley-Walker} N.,  et~al., 2017, \mn@doi [\mnras] {10.1093/mnras/stw2337},
  \href {https://ui.adsabs.harvard.edu/abs/2017MNRAS.464.1146H} {464, 1146}

\bibitem[\protect\citeauthoryear{{Jennison}}{{Jennison}}{1958}]{closurephase}
{Jennison} R.~C.,  1958, \mn@doi [\mnras] {10.1093/mnras/118.3.276}, \href
  {https://ui.adsabs.harvard.edu/abs/1958MNRAS.118..276J} {118, 276}

\bibitem[\protect\citeauthoryear{{Kern}, {Parsons}, {Dillon}, {Lanman},
  {Fagnoni}  \& {de Lera Acedo}}{{Kern} et~al.}{2019}]{systematics1}
{Kern} N.~S.,  {Parsons} A.~R.,  {Dillon} J.~S.,  {Lanman} A.~E.,  {Fagnoni}
  N.,   {de Lera Acedo} E.,  2019, \mn@doi [\apj] {10.3847/1538-4357/ab3e73},
  \href {https://ui.adsabs.harvard.edu/abs/2019ApJ...884..105K} {884, 105}

\bibitem[\protect\citeauthoryear{{Kern} et~al.,}{{Kern}
  et~al.}{2020}]{systematics2}
{Kern} N.~S.,  et~al., 2020, \mn@doi [\apj] {10.3847/1538-4357/ab5e8a}, \href
  {https://ui.adsabs.harvard.edu/abs/2020ApJ...888...70K} {888, 70}

\bibitem[\protect\citeauthoryear{{Kolopanis} et~al.,}{{Kolopanis}
  et~al.}{2019}]{kolopanis2019paper}
{Kolopanis} M.,  et~al., 2019, \mn@doi [\apj] {10.3847/1538-4357/ab3e3a}, \href
  {https://ui.adsabs.harvard.edu/abs/2019ApJ...883..133K} {883, 133}

\bibitem[\protect\citeauthoryear{{Koopmans} et~al.,}{{Koopmans}
  et~al.}{2015}]{skaEoR}
{Koopmans} L.,  et~al., 2015, in Advancing Astrophysics with the Square
  Kilometre Array (AASKA14). p.~1 (\mn@eprint {arXiv} {1505.07568}),
  \mn@doi{10.22323/1.215.0001}

\bibitem[\protect\citeauthoryear{{Lenc} et~al.,}{{Lenc} et~al.}{2017}]{pulsar}
{Lenc} E.,  et~al., 2017, \mn@doi [\pasa] {10.1017/pasa.2017.36}, \href
  {https://ui.adsabs.harvard.edu/abs/2017PASA...34...40L} {34, e040}

\bibitem[\protect\citeauthoryear{{Li} et~al.,}{{Li} et~al.}{2019}]{MWAII}
{Li} W.,  et~al., 2019, \mn@doi [\apj] {10.3847/1538-4357/ab55e4}, \href
  {https://ui.adsabs.harvard.edu/abs/2019ApJ...887..141L} {887, 141}

\bibitem[\protect\citeauthoryear{{Liu}, {Parsons}  \& {Trott}}{{Liu}
  et~al.}{2014}]{eorwindowmath}
{Liu} A.,  {Parsons} A.~R.,   {Trott} C.~M.,  2014, \mn@doi [\prd]
  {10.1103/PhysRevD.90.023018}, \href
  {https://ui.adsabs.harvard.edu/abs/2014PhRvD..90b3018L} {90, 023018}

\bibitem[\protect\citeauthoryear{{Madau}, {Meiksin}  \& {Rees}}{{Madau}
  et~al.}{1997}]{tomography}
{Madau} P.,  {Meiksin} A.,   {Rees} M.~J.,  1997, \mn@doi [\apj]
  {10.1086/303549}, \href
  {https://ui.adsabs.harvard.edu/abs/1997ApJ...475..429M} {475, 429}

\bibitem[\protect\citeauthoryear{Mallat}{Mallat}{2009}]{mallat}
Mallat S.,  2009, A Wavelet Tour of Signal Processing (Third Edition), third
  edition edn.
Academic Press, Boston,
  \mn@doi{https://doi.org/10.1016/B978-0-12-374370-1.X0001-8}

\bibitem[\protect\citeauthoryear{{McKinley} et~al.,}{{McKinley}
  et~al.}{2015}]{FornaxA}
{McKinley} B.,  et~al., 2015, \mn@doi [\mnras] {10.1093/mnras/stu2310}, \href
  {https://ui.adsabs.harvard.edu/abs/2015MNRAS.446.3478M} {446, 3478}

\bibitem[\protect\citeauthoryear{{Mertens} et~al.,}{{Mertens}
  et~al.}{2020}]{mertens2020lofar}
{Mertens} F.~G.,  et~al., 2020, \mn@doi [\mnras] {10.1093/mnras/staa327}, \href
  {https://ui.adsabs.harvard.edu/abs/2020MNRAS.493.1662M} {493, 1662}

\bibitem[\protect\citeauthoryear{{Mesinger}, {Greig}  \& {Sobacchi}}{{Mesinger}
  et~al.}{2016}]{eos}
{Mesinger} A.,  {Greig} B.,   {Sobacchi} E.,  2016, \mn@doi [\mnras]
  {10.1093/mnras/stw831}, \href
  {https://ui.adsabs.harvard.edu/abs/2016MNRAS.459.2342M} {459, 2342}

\bibitem[\protect\citeauthoryear{{Morales} \& {Wyithe}}{{Morales} \&
  {Wyithe}}{2010}]{morales2010reionization}
{Morales} M.~F.,  {Wyithe} J. S.~B.,  2010, \mn@doi [\araa]
  {10.1146/annurev-astro-081309-130936}, \href
  {https://ui.adsabs.harvard.edu/abs/2010ARA&A..48..127M} {48, 127}

\bibitem[\protect\citeauthoryear{{Paciga} et~al.,}{{Paciga}
  et~al.}{2013}]{gmrt}
{Paciga} G.,  et~al., 2013, \mn@doi [\mnras] {10.1093/mnras/stt753}, \href
  {https://ui.adsabs.harvard.edu/abs/2013MNRAS.433..639P} {433, 639}

\bibitem[\protect\citeauthoryear{{Parsons} et~al.,}{{Parsons}
  et~al.}{2010}]{paper}
{Parsons} A.~R.,  et~al., 2010, \mn@doi [\aj] {10.1088/0004-6256/139/4/1468},
  \href {https://ui.adsabs.harvard.edu/abs/2010AJ....139.1468P} {139, 1468}

\bibitem[\protect\citeauthoryear{{Parsons}, {Pober}, {Aguirre}, {Carilli},
  {Jacobs}  \& {Moore}}{{Parsons} et~al.}{2012}]{dspec}
{Parsons} A.~R.,  {Pober} J.~C.,  {Aguirre} J.~E.,  {Carilli} C.~L.,  {Jacobs}
  D.~C.,   {Moore} D.~F.,  2012, \mn@doi [\apj] {10.1088/0004-637X/756/2/165},
  \href {https://ui.adsabs.harvard.edu/abs/2012ApJ...756..165P} {756, 165}

\bibitem[\protect\citeauthoryear{{Patil} et~al.,}{{Patil}
  et~al.}{2017}]{patil2017lofar}
{Patil} A.~H.,  et~al., 2017, \mn@doi [\apj] {10.3847/1538-4357/aa63e7}, \href
  {https://ui.adsabs.harvard.edu/abs/2017ApJ...838...65P} {838, 65}

\bibitem[\protect\citeauthoryear{{Planck Collaboration} et~al.,}{{Planck
  Collaboration} et~al.}{2020}]{planck2018param}
{Planck Collaboration} et~al., 2020, \mn@doi [\aap]
  {10.1051/0004-6361/201833910}, \href
  {https://ui.adsabs.harvard.edu/abs/2020A&A...641A...6P} {641, A6}

\bibitem[\protect\citeauthoryear{{Pritchard} \& {Loeb}}{{Pritchard} \&
  {Loeb}}{2012}]{pritchard2012}
{Pritchard} J.~R.,  {Loeb} A.,  2012, \mn@doi [Reports on Progress in Physics]
  {10.1088/0034-4885/75/8/086901}, \href
  {https://ui.adsabs.harvard.edu/abs/2012RPPh...75h6901P} {75, 086901}

\bibitem[\protect\citeauthoryear{{Rousseeuw} \& {Croux}}{{Rousseeuw} \&
  {Croux}}{1993}]{mad}
{Rousseeuw} P.~J.,  {Croux} C.,  1993, \mn@doi [Journal of the American
  Statistical Association] {10.1080/01621459.1993.10476408}, 88, 1273

\bibitem[\protect\citeauthoryear{{Samuel}, {Nityananda}  \&
  {Thyagarajan}}{{Samuel} et~al.}{2022}]{Samuel+2022}
{Samuel} J.,  {Nityananda} R.,   {Thyagarajan} N.,  2022, \mn@doi [\prl]
  {10.1103/PhysRevLett.128.091101}, \href
  {https://ui.adsabs.harvard.edu/abs/2022PhRvL.128i1101S} {128, 091101}

\bibitem[\protect\citeauthoryear{Shapiro \& Wilk}{Shapiro \&
  Wilk}{1965}]{shapiro-wilk}
Shapiro S.~S.,  Wilk M.~B.,  1965, \mn@doi [Biometrika]
  {10.1093/biomet/52.3-4.591}, 52, 591

\bibitem[\protect\citeauthoryear{{Storer} et~al.,}{{Storer}
  et~al.}{2022}]{antmetrics}
{Storer} D.,  et~al., 2022, \mn@doi [Radio Science] {10.1029/2021RS007376},
  \href {https://ui.adsabs.harvard.edu/abs/2022RaSc...5707376S} {57,
  e2021RS007376}

\bibitem[\protect\citeauthoryear{{Tan} et~al.,}{{Tan} et~al.}{2021}]{errorbar}
{Tan} J.,  et~al., 2021, \mn@doi [\apjs] {10.3847/1538-4365/ac0533}, \href
  {https://ui.adsabs.harvard.edu/abs/2021ApJS..255...26T} {255, 26}

\bibitem[\protect\citeauthoryear{{Thompson}, {Moran}  \& {Swenson}}{{Thompson}
  et~al.}{2017}]{TMS2017}
{Thompson} A.~R.,  {Moran} J.~M.,   {Swenson} George~W. J.,  2017,
  {Interferometry and Synthesis in Radio Astronomy, 3rd Edition}.
Springer, Cham, \mn@doi{10.1007/978-3-319-44431-4}

\bibitem[\protect\citeauthoryear{{Thyagarajan} \& {Carilli}}{{Thyagarajan} \&
  {Carilli}}{2020}]{closuremaths}
{Thyagarajan} N.,  {Carilli} C.~L.,  2020, \mn@doi [\prd]
  {10.1103/PhysRevD.102.022001}, \href
  {https://ui.adsabs.harvard.edu/abs/2020PhRvD.102b2001T} {102, 022001}

\bibitem[\protect\citeauthoryear{{Thyagarajan}, {Carilli}  \&
  {Nikolic}}{{Thyagarajan} et~al.}{2018}]{closure1}
{Thyagarajan} N.,  {Carilli} C.~L.,   {Nikolic} B.,  2018, \mn@doi [\prl]
  {10.1103/PhysRevLett.120.251301}, \href
  {https://ui.adsabs.harvard.edu/abs/2018PhRvL.120y1301T} {120, 251301}

\bibitem[\protect\citeauthoryear{{Thyagarajan} et~al.,}{{Thyagarajan}
  et~al.}{2020}]{closurelimits}
{Thyagarajan} N.,  et~al., 2020, \mn@doi [\prd] {10.1103/PhysRevD.102.022002},
  \href {https://ui.adsabs.harvard.edu/abs/2020PhRvD.102b2002T} {102, 022002}

\bibitem[\protect\citeauthoryear{{Thyagarajan}, {Nityananda}  \&
  {Samuel}}{{Thyagarajan} et~al.}{2022}]{Thyagarajan+2022}
{Thyagarajan} N.,  {Nityananda} R.,   {Samuel} J.,  2022, \mn@doi [\prd]
  {10.1103/PhysRevD.105.043019}, \href
  {https://ui.adsabs.harvard.edu/abs/2022PhRvD.105d3019T} {105, 043019}

\bibitem[\protect\citeauthoryear{{Tingay} et~al.,}{{Tingay} et~al.}{2013}]{mwa}
{Tingay} S.~J.,  et~al., 2013, \mn@doi [\pasa] {10.1017/pasa.2012.007}, \href
  {https://ui.adsabs.harvard.edu/abs/2013PASA...30....7T} {30, e007}

\bibitem[\protect\citeauthoryear{{Trott} et~al.,}{{Trott}
  et~al.}{2020}]{trott2020mwa}
{Trott} C.~M.,  et~al., 2020, \mn@doi [\mnras] {10.1093/mnras/staa414}, \href
  {https://ui.adsabs.harvard.edu/abs/2020MNRAS.493.4711T} {493, 4711}

\bibitem[\protect\citeauthoryear{{Virtanen} et~al.,}{{Virtanen}
  et~al.}{2020}]{scipy}
{Virtanen} P.,  et~al., 2020, \mn@doi [Nature Methods]
  {https://doi.org/10.1038/s41592-019-0686-2}, \href {https://rdcu.be/b08Wh}
  {17, 261}

\bibitem[\protect\citeauthoryear{{Zernike}}{{Zernike}}{1938}]{zernike1938concept}
{Zernike} F.,  1938, \mn@doi [Physica] {10.1016/S0031-8914(38)80203-2}, \href
  {https://ui.adsabs.harvard.edu/abs/1938Phy.....5..785Z} {5, 785}

\bibitem[\protect\citeauthoryear{{Zheng} et~al.,}{{Zheng} et~al.}{2017}]{gsm16}
{Zheng} H.,  et~al., 2017, \mn@doi [\mnras] {10.1093/mnras/stw2525}, \href
  {https://ui.adsabs.harvard.edu/abs/2017MNRAS.464.3486Z} {464, 3486}

\bibitem[\protect\citeauthoryear{{van Cittert}}{{van
  Cittert}}{1934}]{cittert1934wahrscheinliche}
{van Cittert} P.~H.,  1934, \mn@doi [Physica] {10.1016/S0031-8914(34)90026-4},
  \href {https://ui.adsabs.harvard.edu/abs/1934Phy.....1..201V} {1, 201}

\bibitem[\protect\citeauthoryear{{van Haarlem} et~al.,}{{van Haarlem}
  et~al.}{2013}]{lofar}
{van Haarlem} M.~P.,  et~al., 2013, \mn@doi [\aap]
  {10.1051/0004-6361/201220873}, \href
  {https://ui.adsabs.harvard.edu/abs/2013A&A...556A...2V} {556, A2}

\bibitem[\protect\citeauthoryear{{van der Walt}, {Colbert}  \&
  {Varoquaux}}{{van der Walt} et~al.}{2011}]{numpy}
{van der Walt} S.,  {Colbert} S.~C.,   {Varoquaux} G.,  2011, Computing in
  Science Engineering, 13, 22

\makeatother
\end{thebibliography}

%\begin{thebibliography}{99}
%\end{thebibliography}

%%%%%%%%%%%%%%%%%%%%%%%%%%%%%%%%%%%%%%%%%%%%%%%%%%

%%%%%%%%%%%%%%%%% APPENDICES %%%%%%%%%%%%%%%%%%%%%

\appendix
\section{RFI Avoidance in Averaged Closure Phase Data}
\label{sec:rfi}
\begin{figure}
    \centering
    \includegraphics[width=\linewidth]{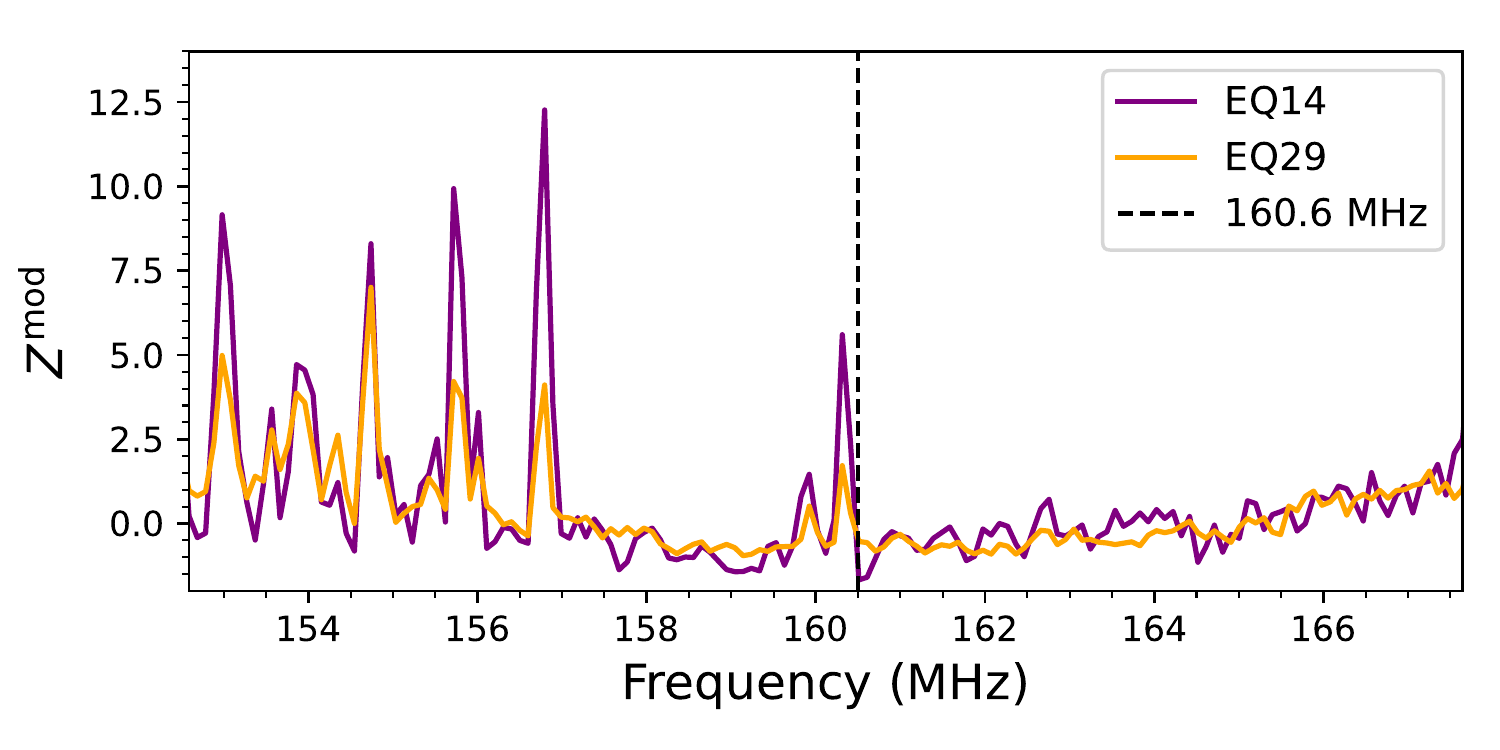}
    \caption{Modified $Z$-score spectrum of time averaged, high-pass filtered closure phase data. EQ14 is shown in purple and EQ29 in orange. The dashed line shows the lower bound of our selected band..}
    \label{fig:rfi2}
\end{figure}
We briefly describe how we selected our frequency band for this analysis. Unlike the visibility processing, our RFI treatment relies on robust averaging rather than flagging and inpainting. However, this strategy is only effective if the RFI is transient and does not repeat on a nightly basis (cf. Section~\ref{sec:avg1}). Hence, before proceeding with the computation of the cross-power spectra, we search for residual RFI in the averaged closure phase data.

First, we average the median-averaged data across triads and polarisation products using inverse variance weights (computed as in Section~\ref{sec:avg2}). The resulting data is two dimensional, consisting of a time and a frequency axis. The aim is to identify features in this spectrogram that, if included, would leak into the high-delay region of the power spectrum. This can be achieved using a wavelet high-pass filter. A complete treatment of wavelets would go beyond the scope of this paper, but we give a brief delineation of the concept. 

Wavelets are functions that are localised in frequency and delay (or traditionally time and frequency) and are generated by scaling and translating a common function called the mother wavelet. A wavelet transform decomposes a signal into a set of wavelets of different scales and shifts. The concept of the wavelet high-pass filter is to approximate the high-delay components of the original signal using only the finest-scale "detail" coefficients. The filter will then be particularly sensitive to features that have a similar shape to the wavelet. Here, we perform a stationary wavelet transform using a Symlet with two vanishing moments \citep{daubechies1988orthonormal} and use the resulting first-level detail coefficients to construct the high-pass filtered data.

The advantage of using wavelets is that they offer sparse representations of a signal, meaning that a signal can be approximated by a small number of wavelets. The smooth components of a signal are represented by large scale wavelets while singularities are represented by highly localised fine scale wavelets. Ringing effects due to discontinuities are highly localised in wavelet approximations. In contrast, Fourier approximations suffer from Gibbs oscillations, which can affect large parts of the approximated signal \citep{mallat}.

To get a picture of the RFI situation across the spectrum of Band~2, we average the absolute values of the high-pass filtered data over time. The spectrum thus obtained will mainly consist of noise fluctuations around zero and peaks corresponding to spectral discontinuities (e.g. RFI). To identify peaks exceeding the noise level, we compute a modified $Z$-score \citep[cf.][]{upperlimits}, defined as

\begin{align}    
    Z^\mathrm{mod}_i &= \frac{\big ||x_i| - \mathrm{med}\{|x|\}\big |}{\sigma_\mathrm{MAD}}\\
    \sigma_\mathrm{MAD} &= 1.4826 \times \mathrm{med}\left\{\big ||x| - \mathrm{med}\{|x|\}\big |\right\},
\end{align}
where $\sigma_\mathrm{MAD}$ is the Median Absolute Deviation (MAD) calculated across frequency, $x_i$ is an element of the averaged high-pass filtered data $x$ and the factor of 1.4826 ensures the equivalence between the standard deviation and the MAD for normally distributed data \citep{mad}. Figure~\ref{fig:rfi2} shows the modified $Z$-score of the high-pass filtered and time-averaged spectrum. From this it is clear that the lower part of Band~2 is significantly contaminated by residual RFI. We choose to set the lower frequency of our band at 160.59\,MHz above which there is little to no significant RFI contamination.

\section{Coherent Time Averaging and Signal Loss}
\label{sec:lstavg}
As explained in Section~\ref{sec:avg1}, the averaging of closure phases in time will lead to a small loss in sensitivity to the cosmological signal. To determine the scale of this loss, we use the EoR model described in Section~\ref{sec:modelling}. At a given LST, we then place GLEAM point sources on the grid defined by the EoR model and weight by the beam response of a HERA dish. Note that it is important to include foregrounds in these simulations since the closure phase is a higher-order quantity. The closure phase will hence consist of cross-terms between the foregrounds and the EoR signal (see Section~\ref{sec:theory}) and the fluctuations due to the latter depend on the foreground structure through Equation~\ref{eq:dphi}. Different foreground structures could therefore cause different degrees of sensitivity-loss under coherent time-averaging.

\begin{figure}
    \centering
    \includegraphics[width=\linewidth]{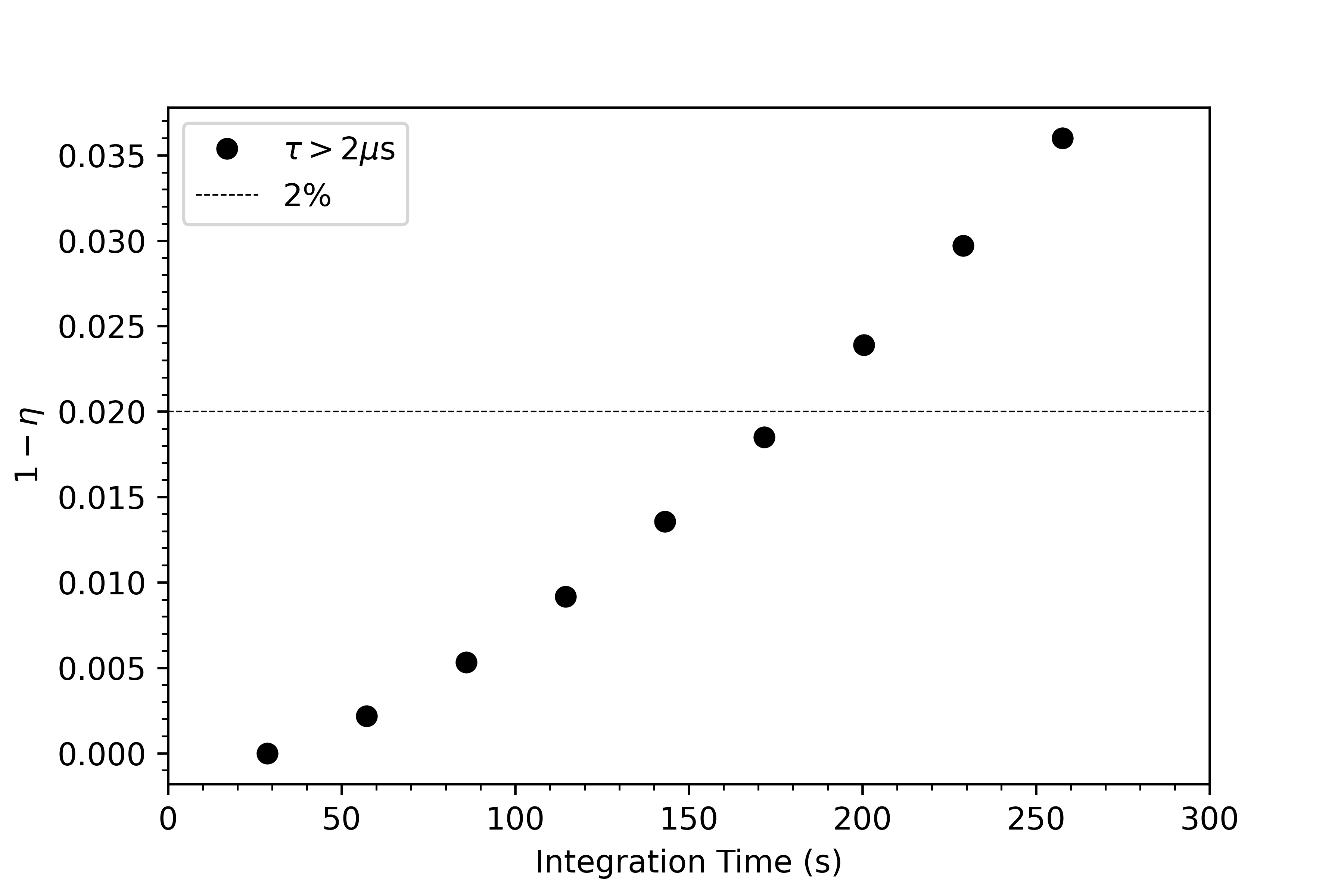}
    \caption{The average fractional loss at $\tau>2\,\mu$s as a function of integration time. Here, we average in intervals of 171.2\,s, which corresponds to a loss smaller than 2\%.}
    \label{fig:lstavg}
\end{figure}

To emulate the change of the apparent sky, we shift the beam with respect to the sky in the interval of a pixel corresponding to $\sim$7' or an observing interval of about 28.6\,s. We compute closure phases and their delay spectra for these sky models. Using these, we can define the fractional loss as:
\begin{equation}
    1 - \eta = \frac{\left< \left| \Psi(t, \tau) \right|^2 \right>_t - \left| \left< \Psi(t, \tau) \right>_t \right|^2}{\left< \left| \Psi(t, \tau) \right|^2 \right>_t},
\end{equation}
where $\eta$ is the factor by which the EoR signal is reduced in the power spectrum and the angular brackets denote an average over a time interval $\Delta t$. We compute fractional losses for 1000 pointings within Field C and average them together. The total fractional loss thus obtained is plotted in Figure~\ref{fig:lstavg} as a function of $\Delta t$. It can be seen that the overall loss at $\Delta t = 171.2$\,s as used in the data analysis is below $2\%$. In a similar analysis \cite{HERAvalidation} find that averaging visibilities over an interval of $240\,$s produces a loss of $\sim1\%$.
The observing interval in our model is greater than the true interval of 10.7\,s, which results in a slight underestimation of the loss. However, the loss at short intervals is considerably lower than the overall loss and should therefore be considered a small effect.

%%%%%%%%%%%%%%%%%%%%%%%%%%%%%%%%%%%%%%%%%%%%%%%%%%

% Don't change these lines
\bsp	% typesetting comment
\label{lastpage}
\end{document}